\pdfoutput=1
\RequirePackage{ifpdf}
\ifpdf 
\documentclass[pdftex]{sigma}
\else
\documentclass{sigma}
\fi

\numberwithin{equation}{section}

\newtheorem{Theorem}{Theorem}[section]
\newtheorem*{Theorem*}{Theorem}

 { \theoremstyle{definition}

\newtheorem{algorithm}[Theorem]{Algorithm}
 }

\usepackage{mathdots}
\usepackage{mathabx}
\DeclareMathOperator{\sech}{sech}
\DeclareMathOperator{\csch}{csch}

\DeclareMathOperator{\tr}{tr}

\begin{document}
\allowdisplaybreaks

\newcommand{\arXivNumber}{1805.12228}

\renewcommand{\PaperNumber}{019}

\FirstPageHeading

\ShortArticleName{Classification of the Orthogonal Separable Webs}

\ArticleName{Classification of the Orthogonal Separable Webs\\ for the Hamilton--Jacobi and Klein--Gordon Equations\\ on 3-Dimensional Minkowski Space}

\Author{Carlos VALERO~$^{\rm a}$ and Raymond G.~MCLENAGHAN~$^{\rm b}$}

\AuthorNameForHeading{C.~Valero and R.G.~McLenaghan}

\Address{$^{\rm a)}$~Department of Mathematics and Statistics, McGill University,\\
\hphantom{$^{\rm a)}$}~Montr\'eal, Qu\'ebec, H3A~0G4, Canada}
\EmailD{\href{mailto:charles.valero@mail.mcgill.ca}{charles.valero@mail.mcgill.ca}}

\Address{$^{\rm b)}$~Department of Applied Mathematics, University of Waterloo,\\
\hphantom{$^{\rm b)}$}~Waterloo, Ontario, N2L~3G1, Canada}
\EmailD{\href{mailto:rgmlena@uwaterloo.ca}{rgmclenaghan@uwaterloo.ca}}

\ArticleDates{Received July 03, 2021, in final form March 02, 2022; Published online March 12, 2022}

\Abstract{We review a new theory of orthogonal separation of variables on pseudo-Rie\-man\-nian spaces of constant zero curvature via concircular tensors and warped products. We then apply this theory to three-dimensional Minkowski space, obtaining an invariant classification of the forty-five orthogonal separable webs modulo the action of the isometry group. The eighty-eight inequivalent coordinate charts adapted to the webs are also determined and listed. We find a number of separable webs which do not appear in previous works in the literature. Further, the method used seems to be more efficient and concise than those employed in earlier works.}


\Classification{53Z05; 70H20; 83A05}

\section{Introduction}\label{sec1}

In this paper we consider the Hamilton--Jacobi equation for geodesics on an $n$-dimensional pseudo-Riemannian manifold $(M,g)$
\begin{gather}\label{eq:Hamilton--Jacobi}
	\frac{1}{2}g^{ij}(q)\frac{\partial W}{\partial q^i}\frac{\partial W}{\partial q^j}= E,
\end{gather}
where $q=\big(q^1,\dots,q^n\big)$ denotes a local coordinate system, $g^{ij}$ the contravariant components of the metric tensor $g$, and $E$ is a non-zero constant.
We also consider the Klein--Gordon equation
\begin{gather}\label{eq:wave}
 \frac{1}{\sqrt{|\det{g}|}} \frac{\partial}{\partial q^i} \left( \sqrt{|\det{g}|} g^{ij} \frac{\partial \varphi}{\partial q^j} \right) +m^2\varphi= 0,
\end{gather}
where $\det{g}$ denotes the determinant of $(g_{ij})$, and $m\neq 0$ is constant.

By additive separability of equation \eqref{eq:Hamilton--Jacobi} with respect to a coordinate system $q$, we mean that the equation admits a {\em complete integral}, i.e., a solution of the form
\begin{gather*}
 W(q,c)=\sum_{i=1}^{n}W_i\big(q^i,c\big),
\end{gather*}
where $c=(c_1,\dots,c_n)$ denotes $n$ constants which satisfy the completeness relation
\begin{gather*}
 \det \left(\frac{\partial^2W}{\partial c_i \partial q^j } \right)\neq0.
\end{gather*}
Product separability of equation \eqref{eq:wave} in a coordinates system $q$ means that the equation possesses a {\em complete separated solution} of the form~\cite{Benenti2002a}
\begin{gather*}
 \varphi(q,c)=\prod_{i=1}^n\varphi_i\big(q^i,c\big),
\end{gather*}
that depend on $2n$ parameters $c=(c_1,\dots,c_{2n})$ which satisfy the completeness relation \cite{Benenti2002a}
\begin{gather*}
 \det \begin{pmatrix} \dfrac{\partial u_i}{\partial c} \\[9pt] \dfrac{\partial v_i}{\partial c} \end{pmatrix} \neq 0, \qquad u_i= \frac{\varphi_i'}{\varphi_i}, \qquad v_i=\frac{\varphi_i''}{\varphi_i}.
\end{gather*}
Separation of variables with respect to coordinates $q$ is said to be {\em orthogonal} if the coordinates $q$ are orthogonal, i.e., if $g_{ij}=0$, for all $i\neq j$. The separation of variables property for a coordinate system $q$ is preserved by any coordinate transformation with diagonal Jacobian. An {\em orthogonal web} on $(M,g)$ is a set of $n$ mutually transversal and orthogonal foliations of dimension $n-1$. A~coordinate system $q$ is {\em adapted} to the web if its leaves are represented locally by $q^i= c^i$, where the $c^i$ are constant parameters. An orthogonal web is said to be {\em separable} if the Hamilton--Jacobi equation is separable in any system of coordinates adapted to the web~\cite{Benenti2002a}. Such a web is called an {\em orthogonal separable web}. It may be shown that if the Klein--Gordon equation is separable in coordinates~$q$, then the Hamilton--Jacobi equation is separable in the same coordinates \cite{Robertson1927}. However, in a space of {\em constant curvature} where the Riemann curvature tensor has the form
\begin{gather*}
 R_{ijkl}=k(g_{ik}g_{jl}-g_{il}g_{jk}),
\end{gather*}
where $k$ is constant, the separability of the Klein--Gordon equation in coordinates $q$ is equivalent to the separability of the Hamilton--Jacobi equation in the same system \cite{Eisenhart1934}. Furthermore, the separable coordinates are necessarily orthogonal \cite{Kalnins1986}. We will say that two orthogonal webs in a pseudo-Riemannian manifold are {\em inequivalent} if they cannot be mapped into each other by an isometry.

The purpose of the present paper is to determine by a new method the inequivalent orthogonal webs on flat 3-dimensional Minkowski space, $\mathbb{E}^3_1$, whose adapted coordinate systems permit additive (resp.\ product) separation for the Hamilton--Jacobi equation for the geodesics (resp.\ Klein--Gordon equation) and to contrast our solution with previously known results. The article is an extension of the work of Rajaratnam, McLenaghan and Valero \cite{Rajaratnam2016} who solved a similar problem for 2-dimensional Minkowski space $\mathbb{E}^2_1$ and de Sitter space $\mathrm{dS}_2$. McLenaghan and Valero have also used this method to obtain a classification of the orthogonal separable webs for 3-dimensional hyperbolic and de Sitter spaces \cite{Valero2019}. A classification of separable webs in these spacetimes could be used to solve boundary value problems whose geometries are adapted to one of these coordinate systems, or to study the integrability of Hamilton--Jacobi or Klein--Gordon equations when coupled to an external field. For example, in \cite{McLenaghan2020}, all separable webs that separate the Hamilton--Jacobi equation for a test charge moving in the electromagnetic field produced by a point particle are classified, and it is therein concluded that only the Cavendish-Coulomb field admits complete separation.

The approach used in this paper is based on the theory of concircular tensors and warped products developed by Rajaratnam~\cite{Rajaratnam2014} and Rajaratnam and McLenaghan \cite{Rajaratnam2014b,Rajaratnam2014a}. It is a~synthesis of the results of Kalnins~\cite{Kalnins1986b}, Crampin~\cite{Crampin2003} and Benenti \cite{Benenti2005b} that has been extended to include general pseudo-Riemannian spaces with application
to pseudo-Riemannian spaces of constant curvature \cite{Rajaratnam2016,Valero2019}. This theory is derived from Eisenhart's \cite{Eisenhart1934} characterization of orthogonal separability by means of valence-two Killing tensors which have simple eigenvalues and orthogonally integrable eigendirections, called {\em characteristic Killing tensors}. We recall that a~{\em Killing tensor} is a symmetric tensor $K_{ij}$ which satisfies the equation \cite{Eisenhart1934}
\begin{gather*}
 \nabla_{i}K_{jk}+\nabla_{j}K_{ki}+\nabla_{k}K_{ij}=0,
\end{gather*}
where $\nabla$ denotes the covariant derivative associated to the Levi-Civita connection of $g$.

This problem has been studied via other methods by different authors including Kalnins \cite{Kalnins1975}, Kalnins and Miller \cite{Kalnins1976a}, Kalnins, Kress and Miller \cite{Kalnins2018}, Hinterleitner \cite{Hinterleitner1996,Hinterleitner1998}, and Horwood and McLenaghan \cite{Horwood2008a}. In \cite{Kalnins1976a} it is shown by a method based on the use of pentaspherical coordinates~\cite{Bocher1894} that the coordinate systems which allow separation have the property that the coordinate surfaces are orthogonal families of confocal quadrics or their limits. The distinct classes of separable coordinates are then classified under the action of the isometry group. In~\cite{Hinterleitner1996,Hinterleitner1998} the coordinate domains and horizons for the coordinate systems of Kalnins and Miller which cannot be reduced to separable coordinate systems in two dimensions are constructed by means of projective plane coordinates. This work adds global considerations to the otherwise local aspects of separable of variables in this case.

In \cite{Horwood2008a} the canonical separable coordinates and their associated webs are found by a purely geometric method which involves the systematic integration of the Killing tensor equations together with the flatness condition in a general orthogonal coordinate system. The transformation to pseudo-Cartesian coordinates is then obtained by the method described in \cite{Horwood2007}. The components of these tensors with respect to pseudo-Cartesian coordinates are obtained from the components in canonical separable coordinates by applying the tensor transformation law. This procedure defines the Killing tensor in question on a coordinate patch of $\mathbb{E}^3_1$. Because the components of the general Killing tensor are polynomial functions in pseudo-Cartesian coordinates, the Killing tensor is defined on all of $\mathbb{E}^3_1$ by analytic continuation. The separable webs are finally obtained as the integral curves of the eigenvector fields of the Killing tensor. Using this procedure Horwood and McLenaghan \cite{Horwood2008b} found thirty-nine orthogonally separable webs and fifty-eight inequivalent metrics in adapted coordinate systems which permit orthogonal separation of variables for the associated Hamilton--Jacobi equation and Klein--Gordon equations. In a subsequent paper Horwood, McLenaghan and Smirnov \cite{Horwood2009} employ the {\em invariant theory of Killing tensors} \cite{Cochran2011,Cochran2017,Horwood2008b, McLenaghan2002,McLenaghan2004,Smirnov2004} to classify the separable webs of $\mathbb{E}^3_1$ by sets of functions of the coefficients of the characteristic Killing tensors which are invariant under the action of the isometry group $E(2,1)$ of $\mathbb{E}^3_1$.

The use of the Rajaratnam et al.\ \cite{Rajaratnam2014b,Rajaratnam2014a} theory to solve this problem has a number of advantages:
\begin{enumerate}\itemsep=0pt
 \item It gives directly the coordinate transformation between the separable coordinates and pseudo-Cartesian coordinates.
 \item It gives directly the expression for the metric tensor in the separable coordinates.
 \item It gives a partially invariant characterization of the separable webs under the appropriate equivalence relation (see Section~\ref{sec2}).
 \item It provides an apparently more compact treatment compared to the other methods described above.
\end{enumerate}

Our calculations yield forty-five orthogonal separable webs which include the thirty-nine found in \cite{Horwood2008a} as well as six additional webs, one of which does not appear in \cite{Hinterleitner1998}. We also note that the number of inequivalent adapted coordinate systems we obtain does not agree with the number given in \cite{Hinterleitner1998}. In this paper we shall provide what appear to be the missing webs in \cite{Horwood2008a} and discuss the discrepancies with the results of \cite{Hinterleitner1998}.

The paper is organized as follows: in Section~\ref{sec2} we review the theory of concircular tensors, their algebraic classification and how certain classes of concircular tensors can be used to construct separable coordinates. In Section~\ref{sec3} the theory of warped products is summarized and an algorithm is given for the construction of reducible separable webs. In both Sections~\ref{sec2} and~\ref{sec3}, we restrict ourselves to the case of constant zero curvature; for a compact review of the theory in the case of constant non-zero curvature, we direct the reader to \cite[Sections~2 and~3]{Valero2019}. In Section~\ref{sec4}, the theory reviewed in Sections~\ref{sec2} and~\ref{sec3} are applied to obtain a complete list of the separable webs and inequivalent adapted coordinate systems on $\mathbb{E}^3_1$. For ease of comparison with previous results we indicate in parentheses how the web is classified in \cite{Horwood2008a}; we also indicate which webs are not found therein. For the irreducible webs, we also give in square brackets the notation used by~\cite{Hinterleitner1998} and~\cite{Kalnins1976a}, again indicating which webs are not found therein. Concluding remarks are given in Section~\ref{sec5}.

We end this section with a word on notation and conventions. We denote the $n$-dimen\-sio\-nal pseudo-Euclidean space with signature $\nu$ by $\mathbb{E}^n_\nu$. By $n$-dimensional Minkowski space, we mean~$\mathbb{E}^n_1$. As $\mathbb{E}^n_\nu$ is a vector space, we make liberal use of the canonical isomorphism between~$\mathbb{E}^n_\nu$ and~$T_p\mathbb{E}^n_\nu$ to identify points and tangent vectors. Accordingly, we denote the metric in~$\mathbb{E}^n_\nu$ by both $g$ and~$\langle \cdot , \cdot \rangle$. Furthermore, as the metric induces an isomorphism between tensor spaces via `raising and lowering indices', we will adopt the standard practice of identifying tensors under this correspondence, using the same symbol regardless of type. It may be convenient to sometimes make this distinction explicit for vectors. In that case, we will write~$v^{\flat}$ for the one-form corresponding to the vector~$v$ via the isomorphism induced by the metric.

Finally, this paper relies heavily on some elementary results regarding the classification of self-adjoint linear operators on $\mathbb{E}^n_\nu$. A review of the relevant material may be found in \cite{Rajaratnam2016}; for convenience, in the appendix we have summarized the main results for the special case of Minkowski space. Any reader not familiar with this subject should look at the appendix before proceeding. The notation and definitions found therein will be used hereafter without comment.

\section[Concircular tensors in E\textasciicircum{}n\_nu]{Concircular tensors in $\boldsymbol{\mathbb{E}^n_\nu}$}\label{sec2}

We recall the basic properties of concircular tensors as they relate to separation of variables. In this section, let $(M,g)$ be a pseudo-Riemannian manifold. A concircular tensor (CT) $L$ on $M$ is a $2$-tensor satisfying
\begin{gather*}
 \nabla_k L_{ij} = \alpha_i g_{jk} + \alpha_j g_{ik}
\end{gather*}
for some $1$-form $\alpha$ on $M$. These tensors are generalization of concircular vectors, which were so-called since they arise as gradients~$\nabla \rho$ of functions $\rho$ such that the conformal scaling~$g \mapsto \rho^2 g$ maps circles to circles; for more information on concircular vectors, see~\cite{Crampin2007}.

We say that $L$ is an \textit{orthogonal}, concircular tensor, if it is pointwise diagonalizable. We say that $L$ is a {\em Benenti tensor} if its eigenfunctions are pointwise simple, i.e., if its eigenvalues at each point are distinct. We call~$L$ an {\em irreducible} concircular tensor, if its eigenfunctions $\lambda_1,\dots,\lambda_n$ are functionally independent. A CT that is not irreducible is called {\em reducible}. One can show that an orthogonal CT is irreducible if and only if none of its eigenfunctions are constant~\cite{Rajaratnam2014a}. We will only need to work with orthogonal CTs in classifying separable webs on spaces of constant curvature. It is therefore of no harm in assuming all CTs to be orthogonal from this point forward.

One can show \cite{Crampin2003} that the general concircular tensor $L$ in $\mathbb{E}^n_\nu$ is given by
\begin{gather*}
 L = A + 2 w \odot r + m r \odot r,
\end{gather*}
where $A$ is a constant symmetric tensor, $w \in \mathbb{E}^n_\nu$, $m \in \mathbb{R}$, $\odot$ is the symmetric tensor product, and $r$ is the \textit{dilatational vector field}, defined in pseudo-Cartesian coordinates $(x^i)$ by $r := x^i \partial_i$.

If $L$ is irreducible, then the distributions orthogonal to its eigenspaces are integrable and define a separable web \cite{Rajaratnam2014a}. In this case, the eigenfunctions of $L$ themselves give separable coordinates adapted to this web. If $L$ is reducible, then there are often multiple separable webs adapted to its eigenspaces; for this case we will need a warped product decomposition of $M$ to determine all the separable coordinates associated with $L$. This will be the subject of the next section.

Therefore, any orthogonal CT induces at least one separable web. It is a remarkable fact that in spaces of constant curvature, {\em all} such separable webs arise in this fashion; see \cite{Rajaratnam2014b,Rajaratnam2014a} for a proof of this theorem. However, it is clear that distinct orthogonal CTs may give rise to separable webs which are related by an isometry. So, to classify separable webs modulo isometry, we must classify orthogonal CTs modulo an appropriate equivalence relation. More precisely, let~$L$ and~$L'$ be two concircular tensors on a pseudo-Riemannian manifold $(M,g)$. We say that~$L$ and~$L'$ are {\em geometrically equivalent} if there exists $a \in \mathbb{R} \setminus \{0 \}$, $b \in \mathbb{R}$, and an isometry $\Lambda \in I(M)$ such that $L' = a \big(\Lambda^{-1}\big)^{\ast} L + bg$. The reason for studying the equivalence classes of CTs under this relation lies in the following result: if $L$ and $L'$ are orthogonal CTs on a connected mani\-fold~$M$, with at least one being non-constant, then their associated separable webs are related by an isometry if and only if they are geometrically equivalent~\cite{Rajaratnam2014a}.

This is a good place to mention some useful results which allow one to connect concircular tensors with the general theory of separation using Killing tensors, as is done in \cite{Horwood2008b, Horwood2008a}. We refer the reader to \cite{Rajaratnam2016} for details on the following construction. If $L$ is a CT, we first define
\begin{gather*}
 K_1 := \tr(L)g - L,
\end{gather*}
where $\tr(L)$ denotes the trace of $L$. $K_1$ is a Killing tensor with the same eigenspaces as~$L$; in particular if~$L$ is Benenti, then $K_1$ is a characteristic Killing tensor called the \textit{Killing--Bertrand--Darboux tensor $($KBDT$)$} associated with~$L$. We may then define inductively, for $2 \leq m \leq n-1$,
\begin{gather*}
 K_m := \frac{1}{m}\tr(K_{m-1}L)g - K_{m-1}L,
\end{gather*}
where $K_{m-1}L$ is a product of endomorphisms. One can show \cite{Benenti2005b} that $\{g,K_1,\dots,K_{n-1}\}$ generate a real $n$-dimensional vector space of Killing tensors which are all diagonalized in the separable coordinates induced by~$L$. This space is called the {\em Killing--St\"ackel algebra} associated with the web; it plays an important role in the Killing tensor approach to separation of variables.

We now quote the following theorem from \cite{Rajaratnam2016} which will be absolutely paramount in our classification of concircular tensors (and hence separable webs) in $\mathbb{E}^3_1$. For simplicity, we state the theorem for the special case of $\mathbb{E}^n_1$, which shall be sufficient for our purposes.

\begin{Theorem}\label{classificationtheorem}
 Any orthogonal concircular tensor $\tilde{L} = \tilde{A} + w \otimes r^{\flat} + r \otimes w^{\flat} + m r \otimes r^{\flat}$ in $\mathbb{E}^n_1$, after a possible change of origin and after passing to a geometrically equivalent concircular tensor $L$, admits precisely one of the following canonical forms:
 \begin{itemize}\itemsep=0pt
 \item Cartesian, if $m = 0$ and $w = 0$, then $L = \tilde{A}$,
 \item central, if $m \neq 0$, then $L = A + r \otimes r^{\flat}$,
 \item non-null axial, if $m = 0$ and $\langle w, w \rangle \neq 0$, then there exists a vector $e_1 \in \operatorname{span}\{ w\}$ such that $L = A + e_1 \otimes r^{\flat} + r \otimes e_1^{\flat}$, $Ae_1 = 0$, and $\langle e_1, e_1 \rangle = \varepsilon = \pm 1$,
 \item null axial, if $m = 0$, $w \neq 0$ and $\langle w, w \rangle = 0$, then there is a skew-normal sequence $($see the appendix for the definition of a skew-normal sequence$)$ $\beta = \{e_1 ,\dots, e_k \}$ with $e_1 \in \operatorname{span}\{ w\}$ and $\langle e_1, e_k \rangle = \varepsilon$, which is $A$-invariant, such that $L = A + e_1 \otimes r^{\flat} + r \otimes e_1^{\flat}$ and the restriction of~$A$ to~$\operatorname{span}\{\beta\}$ is represented by $J_k(0)^{\rm T}$. In Minkowski space, either $k = 2$, in which case $\varepsilon = \pm 1$, or $k = 3$, in which case $\varepsilon = 1$.
 \end{itemize}
\end{Theorem}

In Theorem \ref{classificationtheorem}, $\varepsilon$ is called the {\em sign} of~$L$. For central CTs, the sign is defined to be~$1$, and for Cartesian CTs, the sign is not defined. For a proof of Theorem~\ref{classificationtheorem} in a flat space of arbitrary signature, the reader should consult~\cite{Rajaratnam2014}.

Let $L$ be a CT in its canonical form. Then we let $D$ denote the $A$-invariant subspace spanned by $\big\{w, Aw, A^2w, \dots \big\}$. This subspace is either zero (if $w = 0$), or non-degenerate. We define
\begin{gather*}
 A_c := A|_{D^{\bot}}
\end{gather*}
as well as the following two characteristic polynomials:
\begin{gather*}
 p(z) := \det (zI - L), \qquad B(z) := \det (zI - A_c),
\end{gather*}
where the latter determinant is evaluated in $D^{\bot}$. We therefore have that the classification of CTs within each of the cases listed in Theorem \ref{classificationtheorem}, is reducible to a classification of the metric-Jordan canonical forms for $A_c$.

We also note that in the classification of orthogonal CTs as per Theorem \ref{classificationtheorem}, CTs which differ by multiples of the metric in $D^{\bot}$ induce the same separable webs if $\dim{D} \leq 2$. While not explicitly stated in \cite{Rajaratnam2014}, this result is implied by the proof of Theorem \ref{classificationtheorem} given therein. We shall use this result frequently to simplify our canonical forms for the axial concircular tensors.

We finish this section with some results that will allows us to extract explicit formulae for separable coordinates from an irreducible CT. Recall that an irreducible CT determines a separable web, and its eigenfunctions are separable coordinates adapted to this web; we call these \textit{canonical coordinates} associated with the irreducible CT. Furthermore, one can show that a~non-constant CT $L = A + 2w \odot r + mr\odot r$ in $\mathbb{E}^n_\nu$ (for $\nu \leq 1$) is an irreducible CT if and only if $A_c$ has no multidimensional eigenspaces; equivalently, $L$ is reducible if and only if $A_c$ has a~multidimensional eigenspace \cite{Rajaratnam2014}.

The following technical results are derived in \cite{Rajaratnam2014} and allow one to obtain, given an irreducible CT $L$, the transformation equations between the induced canonical coordinates $\big(u^i\big)$, and the lightcone-Cartesian coordinates $\big(x^i\big)$ in which $(A,g)$ take the prescribed form. First, if $L = A + r \otimes r^{\flat}$ is a central irreducible CT in canonical form, and if $A$ and $g$ take the following forms in coordinates~$\big(x^i\big)$, where the matrices $J_k(\lambda)$ and $S_k$ are defined in Appendix~\ref{appA}:
\begin{gather}\label{Ag}
A = J_k(0)^{\rm T} \oplus \operatorname{diag}(\lambda_{k+1},\dots,\lambda_{n}), \qquad g = \epsilon_0S_k \oplus \operatorname{diag}(\epsilon_{k+1},\dots,\epsilon_{n})
\end{gather}
then, from \cite{Rajaratnam2014}, we have the following equations:
\begin{gather} \label{centralICTa}
 \sum_{i=1}^{l+1} x^i x^{l + 2 - i} = -\frac{\epsilon_0}{l!} \left(\frac{{\rm d}}{{\rm d}z} \right)^l \left( \frac{p(z)}{B_{U^{\bot}}(z)} \right) \bigg|_{z=0},\qquad l = 0,\dots, k-1,
\\
 \label{centralICTb}
 \big(x^i\big)^2 = -\epsilon_i \frac{p(\lambda_i)}{B'(\lambda_i)}, \qquad i = k+1,\dots, n,
\end{gather}
where $U$ is the subspace corresponding to $J_k(0)^{\rm T}$, and $B_{U^{\bot}}(z)$ is the characteristic polynomial of $A$ restricted to $U^{\bot}$. The transformation equations from the canonical coordinates $\big(u^i\big)$ and the coordinates $\big(x^i\big)$ can then be obtained by writing
\begin{gather}\label{p(z)}
p(z) = \prod_{i=1}^{n}\big(z - u^i\big).
\end{gather}
Now let $L = A + e_1 \otimes r^{\flat} + r \otimes e_1^{\flat}$ be an axial irreducible CT in canonical form and suppose that in coordinates $\big(x^i\big)$, we have $A = A_d \oplus A_c$ with $A_d = J_k(0)^{\rm T}$ and $g = \epsilon_0 S_k \oplus g_c$. Then we have from~\cite{Rajaratnam2014} that the characteristic polynomial of $L$ takes the form
\begin{gather} \label{axialICTa}
 p(z) = p_d(z)B(z) + \epsilon_0 (p_c(z) - B(z)),
\end{gather}
where $p_d(z)$ is the characteristic polynomial of $L$ restricted to the subspace corresponding to~$A_d$, and is given explicitly as follows (for $k \geq 2$):
\begin{gather} \label{axialICTb}
 p_d(z) = z^k + \sum_{l=2}^k \sum_{i=1}^{l-1} x^{k+1+i-l} x^{k+1-i} z^{k-l} - 2\epsilon_0 \sum_{i=1}^k x^{k-i+1} z^{k-i}.
\end{gather}
Again, the coordinate transformations from lightcone-Cartesian coordinates $\big(x^i\big)$ to the canonical coordinates $(u^i)$ may be obtained by using the equations~\eqref{axialICTa} and~\eqref{axialICTb}, along with the factorization of $p(z)$, just as with central CTs. We note that in the case $A = A_d$, we of course have $p(z) = p_d(z)$. Furthermore, for $k=1$, it is clear that $p_d(z) = z - 2\epsilon_0 x^1$.

The last result from \cite{Rajaratnam2014} that we quote in this section allows one to quickly write down the expression for the metric in canonical coordinates $(u^i)$ associated with an irreducible CT $L$. For $i = 1,\dots,n$,
\begin{gather}
 \label{ICTmetric}
 g_{ii} = \frac{\varepsilon}{4} \frac{\prod_{j \neq i} \big(u^i - u^j\big)}{\prod_{j = 1}^{n-k} \big(u^i - \lambda_j\big)},
\end{gather}
where $\lambda_{1},\dots,\lambda_{n-k}$ are the roots of~$B(z)$, and $\varepsilon$ is the sign of~$L$ (see the remarks following Theorem~\ref{classificationtheorem}). Of course the separable coordinates $\big(u^i\big)$ are orthogonal, so $g_{ij} = 0$ for $i \neq j$.

\section[Warped products in E\textasciicircum{}n\_nu and reducible concircular tensors]{Warped products in $\mathbb{E}^n_\nu$ and reducible concircular tensors}\label{sec3}

Here we recall warped product decompositions of $\mathbb{E}^n_\nu$ and their importance for constructing separable webs in $\mathbb{E}^n_\nu$. In this section, we quote the relevant results, referring the reader to~\cite{Valero2019} or~\cite{Rajaratnam2014} for details. Recall that given pseudo-Riemannian manifolds $(N_i, g_i)$ for $0 \leq i \leq k$ and smooth positive functions $\rho_j$ for $1 \leq j \leq k$, we define the warped product $N_0 \times_{\rho_1} \times \cdots \times_{\rho_k} N_k$ to be the product manifold equipped with the metric
\begin{gather*}
 g := \pi_0^*g_0 + \sum_{i=1}^{k} \big(\rho_i^2 \circ \pi_0\big) \pi_i^* g_i,
\end{gather*}
where $\pi_i\colon N_0 \times \cdots \times N_k \rightarrow N_i$ is the $i$-th projection. We call $N_0$ the geodesic factor and $N_i$ for $i > 0$ the spherical factors. A {\em warped product decomposition} of a manifold $M$ is a diffeomorphism $\psi\colon N_0 \times_{\rho_1} N_1 \times \cdots \times_{\rho_k} N_k \to M$. We will often speak of warped product decompositions of open subsets of $M$ without much distinction. Moreover, we may at times use the terms warped product and warped product decomposition interchangeably.

Let $L$ be a CT in $M$ and let $\psi\colon N_0 \times_{\rho_1} \cdots \times_{\rho_n} N_k \rightarrow M$ be a warped product decomposition of $M$. We say that $\psi$ is adapted to $L$ if for each $i > 0$ and for all points $p \in N_i$, $\psi_*(T_pN_i)$ is an invariant subspace of $L$, meaning that $L$ acting as an endomorphism maps this subspace into itself. In this case, one can show~\cite{Rajaratnam2014a} that the restriction (via $\psi$) of $L$ to $N_0$ is a Benenti tensor, and therefore induces a separable web on $N_0$ which we may lift to $M$ using $\psi$. In particular, if the restriction of $L$ to $N_0$ is an irreducible CT, then its eigenfunctions yield a set of separable coordinates on $N_0$. By choosing a separable web for each of the spherical factors and lifting them to $M$ via~$\psi$, we hence obtain a separable web in~$M$. Therefore, our method for dealing with a reducible CT $L$ will be to find warped product decompositions of $\mathbb{E}^n_\nu$ adapted to~$L$.

Suppose we are given the following data: a point $\bar{p} \in M$; an orthogonal decomposition $T_{\bar{p}}M = V_0 \obot \cdots \obot V_k$ into non-trivial and non-degenerate subspaces; and $k$ mutually orthogonal and linearly independent vectors $a_1, \dots, a_k \in V_0$. We can without loss of generality assume that $\langle \bar{p}, a_i \rangle = 1$ for all $i$ such that $a_i \neq 0$; we say that this data is in {\em canonical form}. Then one can construct a warped product $N_0 \times_{\rho_1} \times \cdots \times_{\rho_k} N_k$ of $\mathbb{E}^n_\nu$ adapted to $L$, passing through $\bar{p}$, such that $T_{\bar{p}}N_i = V_i$ and $N_i$ is a spherical submanifold of $\mathbb{E}^n_\nu$ having mean curvature normal $a_i$ at $\bar{p}$. A detailed exposition of this construction is given in \cite{Valero2019}, for example. For the classification in Section \ref{sec4}, we shall only need warped products with $2$ factors; we thus recall the relevant results from \cite{Valero2019}.

Let $\psi\colon N_0 \times_{\rho_1} N_1 \rightarrow \mathbb{E}^n_\nu$ be a warped product determined by initial data $(\bar{p}; V_0 \obot V_1 ; a_1)$. If $a_1 = 0$, then the warped product is in fact an ordinary Cartesian product, and
\begin{gather*}
\label{CartesianWP}
 \psi(p_0,p_1) = p_0 + p_1.
\end{gather*}
In this case, the image of $\psi$ is all of $\mathbb{E}^n_\nu$, and the warping function is of course just~$1$. Now assume that $a_1$ is non-null. Define $W_0 := V_0 \cap a_1^{\bot}$, and let $P_0\colon \mathbb{E}^n_\nu \rightarrow W_0$ denote the orthogonal projection. Then~$\psi$ takes the form
\begin{gather*}
 \psi(p_0,p_1) = P_0p_0 + \langle a_1, p_0 \rangle (p_1 - c),
\end{gather*}
where $c := \bar{p} - \langle a_1 , a_1 \rangle^{-1}a_1$. The warping function is given by $\rho(p_0) = \langle p_0, a_1 \rangle$. In this case, if $P_1\colon \mathbb{E}^n_\nu \rightarrow \mathbb{R}a_1 \obot V_1$ denotes the orthogonal projection, we have~\cite{Rajaratnam2014}
\begin{gather}\label{nonnullimage}
 \operatorname{Im}(\psi) = \{p \in \mathbb{E}^n_\nu \, | \, \operatorname{sign} \langle P_1(p), P_1(p) \rangle = \operatorname{sign} \langle a_1, a_1 \rangle \}.
\end{gather}
If we restrict our warped product to a domain such that $N_1$ is connected, then we must impose the extra condition $\langle a_1, P_1(p) \rangle > 0$ in \eqref{nonnullimage}. Now consider the case where $a_1$ is lightlike. Then there is another lightlike vector $b \in V_0$ such that $\langle a_1, b \rangle = 1$. Here, we define $W_0 := V_0 \cap \operatorname{span}\{a_1,b\}^{\bot}$ and $W_1 := V_1$, and let $P_i\colon \mathbb{E}^n_\nu \rightarrow W_i$ for $i=0,1$ denote the orthogonal projection. Then $\psi$ takes the form
\begin{gather*}
 \psi(p_0,p_1) = P_0p_0 + \left(\langle b, p_0 \rangle - \frac{1}{2}\langle a_1, p_0 \rangle \langle P_1p_1, P_1p_1 \rangle \right)a_1 + \langle a_1, p_0 \rangle b + \langle a_1, p_0 \rangle P_1p_1.
\end{gather*}
The warping function is $\rho(p_0) = \langle a_1, p_0 \rangle$ and the image of this warped product is
\begin{gather*}
 \operatorname{Im}(\psi) = \big\{p \in \mathbb{E}^n_\nu \, | \, \langle a_1, p \rangle > 0 \big\}.
\end{gather*}

For the remainder of this section, we restrict ourselves to Euclidean and Minkowski spaces, i.e., $\nu = 0$ and $\nu = 1$ respectively. Let $L = A + 2w \odot r + m r\odot r$ be a reducible CT in $\mathbb{E}^n_\nu$. As $\nu \leq 1$, we have that $L$ is reducible iff it is constant, or if $A_c$ has a multi-dimensional eigenspace (see the remarks following Theorem~\ref{classificationtheorem} for the definition of~$A_c$). If~$L$ is constant, then it is diagonalizable with real eigenspaces, and the Cartesian product of these eigenspaces is in fact the warped product adapted to~$L$. If $L$ is non-constant, then we may use the following algorithm from~\cite{Valero2019}, which yields a warped product decomposition adapted to~$L$.

\begin{algorithm}\label{WP}
 Let $L = A + 2w \odot r + mr \odot r$ be a reducible, non-constant CT, and let $\{E_i\}_i$ be the multidimensional eigenspaces of $A_c$. For each $i$:
 \begin{enumerate}\itemsep=0pt
 \item[(i)] If $E_i$ is non-degenerate, choose a unit vector $a_i \in E_i$ and define $V_i := E_i \cap a_i^{\bot}$.
 \item[(ii)] If $E_i$ is a degenerate subspace, then there is a cycle $v_1,\dots,v_r$ of generalized eigenvectors of $A$, such that $v_r \in E_i$ is lightlike. Let $a_i := v_r$, and define $V_i := E_i \cap v_1^{\bot}$. Note that $V_i$ is non-degenerate, and in $\mathbb{E}^n_1$, we have $r \leq 3$.
 \end{enumerate}
 Define $V_0 := V_1^{\bot} \cap \cdots \cap V_k^{\bot}$, and let $\bar{p} \in \mathbb{E}^n_\nu$ be such that the warped product $\psi\colon N_0 \times_{\rho_1} \cdots \times_{\rho_k} N_k \allowbreak \rightarrow \mathbb{E}^n_\nu$ determined by initial data $(\bar{p}; V_0 \obot \cdots \obot V_k; a_1,\dots,a_k)$ is in canonical form. Then $\psi$ is a~warped product adapted to~$L$.
\end{algorithm}

Let us denote the restriction of $L$ to $N_0$ via $\psi$ by $\tilde{L}$. It is given by
\begin{gather*}
 \tilde{L} = \big(\big(\psi^{-1}\big)_* L\big) \big|_{N_0} = \tilde{A} + 2w \odot \tilde{r} + m\tilde{r} \odot \tilde{r},
\end{gather*}
where $\tilde{A} = A|_{V_0}$ and $\tilde{r}$ is the dilatational vector field in $N_0$. As discussed above, $\tilde{L}$ is Benenti and so induces separable coordinates $\big(u_0^j\big)$ on $N_0$ which we may lift to $\mathbb{E}^n_\nu$ via~$\psi$. We choose separable coordinates $\big(u^j_i\big)$ on the spherical factors $N_i$, and lift all of these to $\mathbb{E}^n_\nu$ with~$\psi$. Then the product coordinates $\big(u_0^j,u_1^j,\dots,u_k^j\big)$ parametrize a separable web on~$\mathbb{E}^n_\nu$.

This finally gives us a procedure for constructing separable webs from reducible CTs in~$\mathbb{E}^n_\nu$. Note that obtaining {\em all} separable webs induced by a reducible CT~$L$ requires knowledge of {\em all} the separable webs in the lower-dimensional spaces that could appear as spherical factors in Algorithm~\ref{WP}. In particular, our classification of the separable webs in~$\mathbb{E}^3_1$ will require knowledge of all separable webs in $\mathbb{E}^2$, $\mathbb{E}^2_1$, $\mathbb{H}^2$ and~dS$_2$. These can be found for example in~\cite{Rajaratnam2016}.

\section[Classification of separable webs on E\textasciicircum{}3\_1]{Classification of separable webs on $\boldsymbol{\mathbb{E}^3_1}$}\label{sec4}

In this section we apply the theory of concircular tensors reviewed in the last two sections to classify all $45$ separable webs on~$\mathbb{E}^3_1$ modulo action of the isometry group. We further determine the inequivalent coordinate charts adapted to the webs; recall that two coordinate charts are said to be equivalent if they can be mapped into each other by an isometry. In the following, we make liberal use of the results presented in the previous two sections, particularly equations \eqref{centralICTa}--\eqref{ICTmetric} for irreducible CTs, and Algorithm~\ref{WP} for reducible CTs.

For the purposes of comparison, beside the name of each separable web below, we indicate in parentheses how the web is classified in~\cite{Horwood2008a} using the invariant theory of Killing tensors; we also indicate which webs are not found therein. For the irreducible webs, we also give the notation used by~\cite{Hinterleitner1998} and~\cite{Kalnins1976a} in square brackets, again indicating which webs are not found therein.

For each separable web below, we give the transformation equations between the separable coordinates and Cartesian coordinates, the form of the metric in the separable coordinates, and the coordinate ranges; if a web has more than one inequivalent region, we give the coordinate charts for each. Note that while we only give the charts for a particular domain in each equivalence class, all other charts of the web can be obtained by isometry (often $x \leftrightarrow y$ or some combination of $t \mapsto -t$, $x \mapsto -x$ and $y \mapsto -y$).

For consistency, we shall adopt the convention of using the same lightcone coordinates $(\eta,\xi)$ throughout this section, defined by
$\eta = x + t$, $\xi = \tfrac{1}{2}(x - t)$.
Notice that $\langle \partial_\eta, \partial_\xi \rangle = 1$. Letting $\xi \mapsto -\xi$ gives lightcone coordinates whose basis vectors have scalar product~$-1$.

\subsection[Cartesian CTs, L=A]{Cartesian CTs, $\boldsymbol{L = A}$}

In the Cartesian case, $L = A$ is constant and diagonalizable with real eigenvalues. If $A$ has distinct eigenvalues, $L$ is Benenti and the induced separable web coincides with the eigenspaces of $A$. We therefore obtain the familiar Cartesian coordinates $(t,x,y)$ on~$\mathbb{E}^3_1$.

If $A$ has only one eigenvalue, then it is a multiple of the metric, say $\lambda g$. Thus, being geometrically equivalent to $A - \lambda g \equiv 0$, it is trivial. The only other cases we have to consider are those where $A$ has a $2$-dimensional eigenspace. There are two possibilities:

{\bf 4.1.i.} $A = J_{-1}(1) \oplus J_1(0) \oplus J_1(0)$.
If $A$ has a spacelike multidimensional eigenspace, then upon passing to a geometric equivalent tensor, we may choose a basis such that $A$ takes the above form; i.e., $A = e_0 \otimes e_0^{\flat}$ for some timelike unit vector $e_0$. It is easy to see that this Cartesian CT is reducible. By the remarks preceding Algorithm~\ref{WP}, we see that a warped product which decomposes $L$ is
\begin{align*}
\begin{split}
\psi\colon \ \mathbb{E}^1_1 \times_1 \mathbb{E}^2 &\rightarrow \mathbb{E}^3_1, \\
							(t e_0, p) &\mapsto t e_0 + p.
\end{split}
\end{align*}
We get a separable web on $\mathbb{E}^3_1$ by taking $t$ in the above map as our coordinate on~$\mathbb{E}^1_1$, and lifting any separable web on~$\mathbb{E}^2$. These are the timelike-cylindrical webs, one of which is the Cartesian web already obtained. The separable webs on~$\mathbb{E}^2$ can be found throughout the literature; see, for instance, \cite[Table~1]{Rajaratnam2016}. We thus obtain the following four separable webs from this CT via the above warped product.

 \textit{$1.$ Cartesian web} (spacelike translational web I)
\begin{gather*}
\begin{cases}
{\rm d}s^2 = -{\rm d}u^2 + {\rm d}v^2 + {\rm d}w^2, \\
t=u, \quad x=v, \quad y=w, \\
-\infty < u < \infty, \quad -\infty < v < \infty, \quad -\infty < w < \infty.
\end{cases}
\end{gather*}

 \textit{$2.$ Timelike-cylindrical polar web} (timelike translational web I)
\begin{gather*}
\begin{cases}
{\rm d}s^2 = -{\rm d}u^2 + {\rm d}v^2 + v^2{\rm d}w^2, \\
t=u, \quad x=v \cos{w}, \quad y=v \sin{w}, \\
-\infty < u < \infty, \quad 0 < v < \infty, \quad 0 < w < 2\pi.
\end{cases}
\end{gather*}

\textit{$3.$ Timelike-cylindrical elliptic web}\ \ (timelike translational web III)
\begin{gather*}
\begin{cases}
{\rm d}s^2 = -{\rm d}u^2 + a^2\big(\cosh^2{v} - \cos^2{w}\big)\big({\rm d}v^2 + {\rm d}w^2\big), \\
t=u, \quad x=a\cosh{v}\cos{w}, \quad y=a\sinh{v}\sin{w}, \\
-\infty < u < \infty, \quad 0 < v < \infty, \quad 0 < w < 2\pi, \quad a > 0.
\end{cases}
\end{gather*}

\textit{$4.$ Timelike-cylindrical parabolic web} (timelike translational web II)
\begin{gather*}
\begin{cases}
{\rm d}s^2 = -{\rm d}u^2 + \big(v^2 + w^2\big)\big({\rm d}v^2 + {\rm d}w^2\big), \\
t=u, \quad x=\frac{1}{2}\big(v^2 - w^2\big), \quad y=vw, \\
-\infty < u < \infty, \quad 0 < v < \infty, \quad -\infty < w < \infty.
\end{cases}
\end{gather*}

\textbf{4.1.ii.} $A = J_{-1}(0) \oplus J_1(0) \oplus J_1(1)$.
If $A$ has a multidimensional Lorentzian eigenspace, then using geometric equivalence, we may choose a basis such that~$A$ takes the above form; i.e., $A = e_1 \otimes e_1^{\flat}$ for some spacelike unit vector~$e_1$. This CT is reducible, and by Algorithm~\ref{WP}, we see that a warped product which decomposes $L$ is
\begin{align*}
\psi \colon \ \mathbb{E}^1 \times_1 \mathbb{E}^2_1 &\rightarrow \mathbb{E}^3_1, \\
							(y e_1, p) &\mapsto p + y e_1.
\end{align*}
We obtain a separable web on $\mathbb{E}^3_1$ by taking $y$ in the above map as our coordinate on~$\mathbb{E}^1$, and choosing any separable web on~$\mathbb{E}^2_1$. These are the spacelike-cylindrical webs, one of which is the Cartesian web already obtained. The separable webs on~$\mathbb{E}^2_1$ may be found in \cite[Table~2]{Rajaratnam2016}, for instance. Hence we obtain the following~$9$ (omitting Cartesian coordinates) separable webs from the above warped product.

\textit{$5.$ Spacelike-cylindrical Rindler web} (spacelike translational web II),

\noindent for $-t^2 + x^2 < 0$
\begin{gather*}
\begin{cases}
{\rm d}s^2 = -{\rm d}u^2 + u^2{\rm d}v^2 + {\rm d}w^2, \\
t = u \cosh{v}, \quad x = u \sinh{v}, \quad y=w, \\
0< u < \infty, \quad -\infty < v < \infty, \quad -\infty < w < \infty;
\end{cases}
\end{gather*}
for $-t^2 + x^2 > 0$
\begin{gather*}
\begin{cases}
{\rm d}s^2 = {\rm d}u^2 - u^2{\rm d}v^2 + {\rm d}w^2, \\
t = u \sinh{v}, \quad x = u \cosh{v}, \quad y=w, \\
0< u < \infty, \quad -\infty < v < \infty, \quad -\infty < w < \infty.
\end{cases}
\end{gather*}

\textit{$6.$ Spacelike-cylindrical elliptic web I} (spacelike translational web VI)
\begin{gather*}
\begin{cases}
{\rm d}s^2 = a^2\big(\cosh^2{u} + \sinh^2{v}\big)\big({\rm d}u^2 - {\rm d}v^2\big) + {\rm d}w^2, \\
t = a \cosh{u} \sinh{v}, \quad x = a \cosh{v} \sinh{u}, \quad y=w,\\
0< u < \infty, \quad 0 < v < \infty, \quad -\infty < w < \infty, \quad a>0.
\end{cases}
\end{gather*}

\textit{$7.$ Spacelike-cylindrical elliptic web II} (spacelike translational web VII),

\noindent for $|t| - |x| > a$
\begin{gather*}
\begin{cases}
 {\rm d}s^2 = a^2\big(\cosh^2{v} - \cosh^2{u}\big)\big({\rm d}u^2 - {\rm d}v^2\big) + {\rm d}w^2, \\
 t = a \cosh{u} \cosh{v}, \quad x = a \sinh{v} \sinh{u}, \quad y=w,\\
 0< u < v < \infty, \quad -\infty < w < \infty, \quad a > 0;
\end{cases}
\end{gather*}
for $|t| - |x| < - a$
\begin{gather*}
\begin{cases}
{\rm d}s^2 = a^2\big(\cosh^2{u} - \cosh^2{v}\big)\big({\rm d}u^2 - {\rm d}v^2\big) + {\rm d}w^2, \\
t = a \sinh{u} \sinh{v}, \quad x = a \cosh{v} \cosh{u}, \quad y=w,\\
0< v < u < \infty, \quad -\infty < w < \infty, \quad a > 0;
\end{cases}
\end{gather*}
for $|t| + |x| < a$
\begin{gather*}
\begin{cases}
{\rm d}s^2 = a^2\big(\cos^2{u} - \cos^2{v}\big)\big({\rm d}u^2 - {\rm d}v^2\big) + {\rm d}w^2, \\
t = a \cos{u} \cos{v}, \quad x = a \sin{v} \sin{u}, \quad y=w,\\
0< v < u < \frac{\pi}{2}, \quad -\infty < w < \infty, \quad a>0.
\end{cases}
\end{gather*}

\textit{$8.$ Spacelike-cylindrical complex elliptic web} (spacelike translational web VIII)
\begin{gather*}
\begin{cases}
{\rm d}s^2 = a^2(\sinh{2u} + \sinh{2v})\big({\rm d}u^2 - {\rm d}v^2\big) + {\rm d}w^2, \\
t - x = a \cosh{(u + v)}, \quad t+x = a \sinh{(u - v)}, \quad y=w, \\
0< |v| < u < \infty, \quad -\infty < w < \infty, \quad a>0.
\end{cases}
\end{gather*}

\textit{$9.$ Spacelike-cylindrical null elliptic web I} (spacelike translational web IX)
\begin{gather*}
\begin{cases}
{\rm d}s^2 = \big({\rm e}^{2u} + {\rm e}^{2v}\big)\big({\rm d}u^2 - {\rm d}v^2\big) + {\rm d}w^2, \\
t-x = {\rm e}^{u+v}, \quad t+x = 2 \sinh{(u-v)}, \quad y=w, \\
-\infty < u < \infty, \quad -\infty < v < \infty, \quad -\infty < w < \infty.
\end{cases}
\end{gather*}

\textit{$10.$ Spacelike-cylindrical null elliptic web II} (spacelike translational web X),

\noindent for $-t^2 + x^2 > |t - x|$
\begin{gather*}
\begin{cases}
{\rm d}s^2 = \big({\rm e}^{2u} - {\rm e}^{2v}\big)({\rm d}u^2 - {\rm d}v^2) + {\rm d}w^2, \\
t - x = -{\rm e}^{u+v}, \quad t + x = 2 \cosh{(u-v)}, \quad y=w, \\
-\infty < v < u < \infty, \quad -\infty < w < \infty;
\end{cases}
\end{gather*}

\noindent for $-t^2 + x^2 < -|t - x|$
\begin{gather*}
\begin{cases}
{\rm d}s^2 = \big({\rm e}^{2v} - {\rm e}^{2u}\big)\big({\rm d}u^2 - {\rm d}v^2\big) + {\rm d}w^2, \\
t - x = {\rm e}^{u+v}, \quad t + x = 2 \cosh{(u-v)}, \quad y=w, \\
-\infty < u < v < \infty, \quad -\infty < w < \infty.
\end{cases}
\end{gather*}

\textit{$11.$ Spacelike-cylindrical timelike parabolic web} (spacelike translational web III)
\begin{gather*}
\begin{cases}
{\rm d}s^2 = \big(u^2 - v^2\big)\big({-}{\rm d}u^2+ {\rm d}v^2\big) + {\rm d}w^2, \\
t = \tfrac{1}{2}\big(u^2 + v^2\big), \quad x = uv, \quad y=w, \\
0 < v < u < \infty, \quad -\infty < w < \infty.
\end{cases}
\end{gather*}

\textit{$12.$ Spacelike-cylindrical spacelike parabolic web} (spacelike translational web IV)
\begin{gather*}
\begin{cases}
{\rm d}s^2 = \big(u^2 - v^2\big)\big({\rm d}u^2 - {\rm d}v^2\big) + {\rm d}w^2, \\
t = uv, \quad x = \frac{1}{2}\big(u^2 + v^2\big), \quad y=w ,\\
0 < v < u < \infty, \quad -\infty < w < \infty.
\end{cases}
\end{gather*}

\textit{$13.$ Spacelike-cylindrical null parabolic web} (spacelike translational web V)
\begin{gather*}
\begin{cases}
{\rm d}s^2 = (u - v)\big({-}{\rm d}u^2 + {\rm d}v^2\big) + {\rm d}w^2, \\
t + x = u + v, \quad t - x = -\tfrac{1}{2}(u - v)^2, \quad y=w, \\
0 < |v| < u < \infty, \quad -\infty < w < \infty.
\end{cases}
\end{gather*}

\subsection[Central CTs, L=A+r otimes r\textasciicircum{}\{flat\}]{Central CTs, $\boldsymbol{L = A + r \otimes r^{\flat}}$}

Since all central CTs have the this canonical form, we need only classify the inequivalent canonical forms for $A$. Since $A_c = A$ in this case, a central CT $L = A + r \otimes r^{\flat}$ is reducible if and only if $A$ has a multidimensional eigenspace.

\textbf{4.2.i.} $A = 0$.
In this case, $L = r \otimes r^{\flat}$ is reducible, and hence Algorithm~\ref{WP} must be used to construct a warped product which yields the separable webs induced by $L$. This decomposition depends on the point $\bar{p}$ with which we construct our warped product (see Algorithm~\ref{WP}).

For a timelike unit vector $e_0$, Algorithm~\ref{WP} gives the following warped product which decomposes $L$ in the connected timelike region containing $e_0$:
\begin{align*}
\psi \colon \ N_0 \times_\rho \mathbb{H}^2 &\rightarrow \mathbb{E}^3_1, \\
							(-u e_0, p) &\mapsto u p,
\end{align*}
where $N_0 = \big\{{-}ue_0 \in \mathbb{E}^3_1 \, | \, u > 0 \big\}$, and $\rho(-ue_0) = u$. For a spacelike unit vector~$e_1$, Algorithm~\ref{WP} gives the following warped product in the spacelike region:
\begin{align*}
\psi \colon \ N_0 \times_\rho \mathrm{dS}_2 &\rightarrow \mathbb{E}^3_1, \\
							(w e_1, p) &\mapsto w p,
\end{align*}
where $N_0 = \big\{we_1 \in \mathbb{E}^3_1 \, | \, w > 0 \big\}$, and $\rho(we_1) = w$. We obtain an induced separable web in the timelike (resp. spacelike) region by taking the coordinate~$u$ (resp.~$w$) on the geodesic factor, and lifting any separable web on~$\mathbb{H}^2$ (resp.~$\mathrm{dS}_2$). We hence construct a separable web on~$\mathbb{E}^3_1$ by choosing a separable web in the timelike regions, and a corresponding web in the spacelike region (i.e., the webs in each region should correspond to the same characteristic Killing tensor). The separable webs and coordinate systems on $\mathrm{dS}_2$ may be found in~\cite[Table~3]{Rajaratnam2016}, while those for $\mathbb{H}^2$ may be found in, say, \cite{Olevsky1950}~or~\cite{Bruce2001}. We so obtain the following nine dilatationally invariant webs, in which $K(a)$ denotes the complete elliptic integral of the first kind with elliptic modulus~$a$, for $0 < a < 1$.

\textit{$14.$ Dilatational elliptic web I} (not in~\cite{Horwood2008a}),

\noindent for $-t^2 + x^2 + y^2 > 0$
\begin{gather*}
\begin{cases}
{\rm d}s^2 = w^2\big(\operatorname{dc}^2(u;a) - a^2 \operatorname{sn}^2(v;a)\big)\big({-}{\rm d}u^2 + {\rm d}v^2\big) + {\rm d}w^2, \\
t = w \operatorname{sc}(u;a) \operatorname{dn}(v;a), \quad x= w \operatorname{nc}(u;a) \operatorname{cn}(v;a), \quad y = w \operatorname{dc}(u;a) \operatorname{sn}(v;a), \\
0 < u < K(a), \quad 0 < v < K(a), \quad 0 < w < \infty, \quad 0 < a < 1;
\end{cases}
\end{gather*}
for $-t^2 + x^2 + y^2 < 0$
\begin{gather*}
\begin{cases}
{\rm d}s^2 = -{\rm d}u^2 + u^2\big(a^2\operatorname{cd}^2(v;a) + \operatorname{cs}^2(w;b)\big)\big({\rm d}v^2 + {\rm d}w^2\big), \\
t = u \operatorname{nd}(v;a) \operatorname{ns}(w;b), \quad x = u \operatorname{sd}(v;a) \operatorname{ds}(w;b), \quad y = u \operatorname{cd}(v;a) \operatorname{cs}(w;b), \\
0 < u < \infty, \quad\! 0 < v < K(a), \quad\! 0 < w < K(b), \quad\! 0 < a < 1, \quad\! 0 < b < 1, \quad\! a^2 + b^2 = 1.
\end{cases}
\end{gather*}

\textit{$15.$ Dilatational elliptic web II} (dilatational web IV),

\noindent for $-t^2 + x^2 + y^2 > 0$, $a|t| - |x| > b\sqrt{-t^2 + x^2 + y^2}$
\begin{gather*}
\begin{cases}
{\rm d}s^2 = w^2\big(\operatorname{dc}^2(u;a) - \operatorname{dc}^2(v;a)\big)\big({-}{\rm d}u^2 + {\rm d}v^2\big) + {\rm d}w^2, \\
t = \frac{wb}{a} \operatorname{nc}(u;a) \operatorname{nc}(v;a), \quad x= wb \operatorname{sc}(u;a) \operatorname{sc}(v;a), \quad y = \frac{w}{a} \operatorname{dc}(u;a) \operatorname{dc}(v;a), \\
0 < v < u < K(a), \quad 0 < w < \infty, \quad 0 < a < 1, \quad 0 < b < 1, \quad a^2 + b^2 = 1;
\end{cases}
\end{gather*}
for $-t^2 + x^2 + y^2 > 0$, $a|t| + |x| < b\sqrt{-t^2 + x^2 + y^2}$
\begin{gather*}
\begin{cases}
{\rm d}s^2 = w^2a^2\big(\operatorname{nd}^2(u;b) - \operatorname{nd}^2(v;b)\big)\big({\rm d}u^2 - {\rm d}v^2\big) + {\rm d}w^2, \\
t = w a b \operatorname{sd}(u;b) \operatorname{sd}(v;b), \quad x= w b \operatorname{cd}(u;b) \operatorname{cd}(v;b), \quad y = w a \operatorname{nd}(u;b) \operatorname{nd}(v;b), \\
0 < v < u < K(b), \quad 0 < w < \infty, \quad 0 < a < 1, \quad 0 < b < 1, \quad a^2 + b^2 = 1;
\end{cases}
\end{gather*}
for $-t^2 + x^2 + y^2 < 0$
\begin{gather*}
\begin{cases}
{\rm d}s^2 = -{\rm d}u^2 + u^2(\operatorname{dc}^2(v;a) + a^2\operatorname{sc}^2(w;b))\big({\rm d}v^2 + {\rm d}w^2\big), \\
t = u \operatorname{nc}(v;a) \operatorname{nc}(w;b), \quad x = u \operatorname{sc}(v;a) \operatorname{dc}(w;b), \quad y = u \operatorname{dc}(v;a) \operatorname{sc}(w;b), \\
0 < u < \infty, \quad\! 0 < v < K(a), \quad\! 0 < w < K(b), \quad\! 0 < a < 1, \quad\! 0 < b < 1, \quad\! a^2 + b^2 = 1.
\end{cases}
\end{gather*}

\textit{$16.$ Spherical web I} (timelike rotational web I),

\noindent for $-t^2 + x^2 + y^2 > 0$
\begin{gather*}
\begin{cases}
{\rm d}s^2 = w^2\big({-}{\rm d}u^2 + \cosh^2{u} {\rm d}v^2\big) + {\rm d}w^2, \\
t = w \sinh{u}, \quad x= w \cosh{u} \cos{v}, \quad y = w \cosh{u} \sin{v}, \\
-\infty < u < \infty, \quad 0 < v < 2\pi, \quad 0 < w < \infty;
\end{cases}
\end{gather*}
for $-t^2 + x^2 + y^2 < 0$
\begin{gather*}
\begin{cases}
{\rm d}s^2 = -{\rm d}u^2 + u^2\big({\rm d}v^2 + \sinh^2{v}{\rm d}w^2\big),\\
t = u \cosh{v}, \quad x= u \sinh{v} \cos{w}, \quad y = u \sinh{v} \sin{w}, \\
0 < u < \infty, \quad 0 < v < \infty, \quad 0 < w < 2\pi.
\end{cases}
\end{gather*}

{\samepage \textit{$17.$ Spherical web II} (spacelike rotational web I),\\
for $-t^2 + x^2 + y^2 > 0$, $-t^2 + x^2 > 0$
\begin{gather*}
\begin{cases}
{\rm d}s^2 = w^2\big({\rm d}u^2 - \sin^2{u}{\rm d}v^2\big) + {\rm d}w^2, \\
t = w \sin{u} \sinh{v}, \quad x= w \sin{u} \cosh{v}, \quad y = w \cos{u}, \\
0 < u < \pi, \quad -\infty < v < \infty, \quad 0 < w < \infty;
\end{cases}
\end{gather*}}

\noindent
for $-t^2 + x^2 + y^2 > 0$, $-t^2 + x^2 < 0$
\begin{gather*}
\begin{cases}
{\rm d}s^2 = w^2\big({-}{\rm d}u^2 + \sinh^2{u}{\rm d}v^2\big) + {\rm d}w^2, \\
t = w \sinh{u} \cosh{v}, \quad x= w \sinh{u} \sinh{v}, \quad y = w \cosh{u}, \\
0 < u < \infty, \quad -\infty < v < \infty, \quad 0 < w < \infty;
\end{cases}
\end{gather*}
for $-t^2 + x^2 + y^2 < 0$
\begin{gather*}
\begin{cases}
{\rm d}s^2 = -{\rm d}u^2 + u^2\big({\rm d}v^2 + \cosh^2{v}{\rm d}w^2\big), \\
t = u \cosh{v} \cosh{w}, \quad x = u \cosh{v} \sinh{w}, \quad y = u \sinh{v}, \\
0 < u < \infty, \quad -\infty < v < \infty, \quad -\infty < w < \infty.
\end{cases}
\end{gather*}

\textit{$18.$ Dilatational complex elliptic web} (dilatational web V),

\noindent for $-t^2 + x^2 + y^2 > 0$
\begin{gather*}
\begin{cases}
{\rm d}s^2 = w^2\big(\operatorname{sn}^2(u;a)\operatorname{dc}^2(u;a) - \operatorname{sn}^2(v;a)\operatorname{dc}^2(v;a)\big)\big({-}{\rm d}u^2 + {\rm d}v^2\big) + {\rm d}w^2, \\
t^2 + x^2 = \dfrac{2w^2\operatorname{dn}(2u;a) \operatorname{dn}(2v;a)}{ab(1+\operatorname{cn}(2u;a))(1+\operatorname{cn}(2v;a))}, \\
 -t^2 + x^2 = \dfrac{2w^2(\operatorname{cn}(2u;a) + \operatorname{cn}(2v;a))}{(1 + \operatorname{cn}(2u;a))(1 + \operatorname{cn}(2v;a))}, \\
y = \operatorname{sn}(u;a) \operatorname{dc}(u;a)\operatorname{sn}(v;a) \operatorname{dc}(v;a), \\
0 < v < u < K(a), \quad 0 < w < \infty, \quad 0 < a < 1, \quad 0 < b < 1, \quad a^2 + b^2 = 1.
\end{cases}
\end{gather*}
for $-t^2 + x^2 + y^2 < 0$
\begin{gather*}
\begin{cases}
{\rm d}s^2 = -{\rm d}u^2 + u^2\big(\operatorname{sn}^2(v;a)\operatorname{dc}^2(v;a) + \operatorname{sn}^2(w;b)\operatorname{dc}^2(w;b)\big)\big({\rm d}v^2 + {\rm d}w^2\big), \\
t^2 + x^2 = \dfrac{2u^2\operatorname{dn}(2v;a) \operatorname{dn}(2w;b)}{ab(1+\operatorname{cn}(2v;a))(1+\operatorname{cn}(2w;b))}, \\
 t^2 - x^2 = \dfrac{2u^2(1+ \operatorname{cn}(2v;a)\operatorname{cn}(2w;b))}{(1 + \operatorname{cn}(2v;a))(1 + \operatorname{cn}(2w;b))}, \\
y = u \operatorname{sn}(v;a) \operatorname{dc}(v;a)\operatorname{sn}(w;b) \operatorname{dc}(w;b), \\
0 < u < \infty, \quad\! 0 < v < K(a), \quad\! 0 < w < K(b), \quad\! 0 < a < 1, \quad\! 0 < b < 1, \quad\! a^2 + b^2 = 1.
\end{cases}
\end{gather*}

\textit{$19.$ Dilatational null elliptic web I} (dilatational web II),

\noindent for $-t^2 + x^2 + y^2 > 0$
\begin{gather*}
\begin{cases}
{\rm d}s^2 = w^2\big(\sech^2{u} + \csch^2{v}\big)\big({\rm d}u^2 - {\rm d}v^2\big) + {\rm d}w^2, \\
t + x = w \sech{u} \csch{v}, \quad x - t = w \cosh{u} \sinh{v} \big(1 - \tanh^2{u} \coth^2{v}\big), \\
 y = w \tanh{u} \coth{v}, \\
0 < u < \infty, \quad 0 < v < \infty, \quad 0 < w < \infty;
\end{cases}
\end{gather*}
for $-t^2 + x^2 + y^2 < 0$
\begin{gather*}
\begin{cases}
{\rm d}s^2 = -{\rm d}u^2 + u^2\big(\sec^2{v} - \sech^2{w}\big)\big({\rm d}v^2 + {\rm d}w^2\big), \\
t + x = u \sec{v} \sech{w}, \quad t - x = u \cos{v} \cosh{w} \big(1 + \tan^2{v} \tanh^2{w}\big), \\
y = u \tan{v} \tanh{w}, \\
0 < u < \infty, \quad 0 < v < \frac{\pi}{2}, \quad 0 < w < \infty.
\end{cases}
\end{gather*}

\textit{$20.$ Dilatational null elliptic web II} (dilatational web III),

\noindent for $-t^2 + x^2 + y^2 > 0$, $|x| > \sqrt{-t^2 + x^2 + y^2}$, $tx > 0$
\begin{gather*}
\begin{cases}
{\rm d}s^2 = w^2\big(\sec^2{u} - \sec^2{v}\big)\big({-}{\rm d}u^2 + {\rm d}v^2\big) + {\rm d}w^2, \\
t + x = w \sec{u} \sec{v}, \quad t - x = -w \cos{u} \cos{v} \big(1 - \tan^2{u} \tan^2{v}\big), \\
y = w \tan{u} \tan{v}, \\
0 < v < u < \frac{\pi}{2}, \quad 0 < w < \infty;
\end{cases}
\end{gather*}
for $-t^2 + x^2 + y^2 > 0$, $|x| > \sqrt{-t^2 + x^2 + y^2}$, $tx < 0$, $|y| > \sqrt{-t^2 + x^2 + y^2}$
\begin{gather*}
\begin{cases}
{\rm d}s^2 = w^2\big(\csch^2{v} - \csch^2{u}\big)\big({\rm d}u^2 - {\rm d}v^2\big) + {\rm d}w^2, \\
t + x = w \csch{u} \csch{v}, \quad t - x = - w \sinh{u} \sinh{v} \big(1 - \coth^2{u} \coth^2{v}\big), \\
y = w \coth{u} \coth{v}, \\
0 < v < u < \infty, \quad 0 < w < \infty;
\end{cases}
\end{gather*}
for $-t^2 + x^2 + y^2 > 0$, $|x| > \sqrt{-t^2 + x^2 + y^2}$, $tx < 0$, $|y| < \sqrt{-t^2 + x^2 + y^2}$
\begin{gather*}
\begin{cases}
{\rm d}s^2 = w^2\big(\sech^2{u} - \sech^2{v}\big)\big({\rm d}u^2 - {\rm d}v^2\big) + {\rm d}w^2, \\
t + x = w \sech{u} \sech{v}, \quad t - x = - w \cosh{u} \cosh{v} \big(1 - \tanh^2{u} \tanh^2{v}\big), \\
y = w \tanh{u} \tanh{v}, \\
0 < u < v < \infty, \quad 0 < w < \infty;
\end{cases}
\end{gather*}
for $-t^2 + x^2 + y^2 < 0$
\begin{gather*}
\begin{cases}
{\rm d}s^2 = -{\rm d}u^2 + u^2\big(\csch^2v + \sec^2w\big)\big({\rm d}v^2 + {\rm d}w^2\big), \\
t + x = u \csch{v} \sec{w}, \quad t - x = u \sinh{v} \cos{w}\big(1 + \coth^2v \tan^2w\big), \\
 y = u \coth{v} \tan{w}, \\
0 < u < \infty, \quad 0 < v < \infty, \quad 0 < w < \frac{\pi}{2}.
\end{cases}
\end{gather*}

\textit{$21.$ Null spherical web} (null rotational web I),

\noindent for $-t^2 + x^2 + y^2 > 0$
\begin{gather*}
\begin{cases}
{\rm d}s^2 = w^2\big({-}{\rm d}u^2 + {\rm e}^{2u}{\rm d}v^2\big) + {\rm d}w^2, \\
t+x = w\big({\rm e}^{-u} - v^2{\rm e}^u\big), \quad t-x = -w{\rm e}^u, \quad y = wv{\rm e}^u, \\
-\infty < u < \infty, \quad -\infty < v < \infty, \quad 0 < w < \infty;
\end{cases}
\end{gather*}
for $-t^2 + x^2 + y^2 < 0$
\begin{gather*}
\begin{cases}
{\rm d}s^2 = -{\rm d}u^2 + u^2\big({\rm d}v^2 + {\rm e}^{2v}{\rm d}w^2\big), \\
t+x = u{\rm e}^v, \quad t-x = u\big({\rm e}^{-v} + w^2{\rm e}^v\big), \quad y = uw{\rm e}^v, \\
0 < u < \infty, \quad -\infty < v < \infty, \quad -\infty < w < \infty.
\end{cases}
\end{gather*}

\textit{$22.$ Dilatational null elliptic web III} (dilatational web I),

\noindent for $-t^2 + x^2 + y^2 > 0$
\begin{gather*}
\begin{cases}
{\rm d}s^2 = w^2\left(\dfrac{1}{u^2} - \dfrac{1}{v^2}\right)\big({-}{\rm d}u^2 + {\rm d}v^2\big) + {\rm d}w^2, \\
t+x = \dfrac{w}{uv}, \quad t-x = \dfrac{w\big(u^2 - v^2\big)^2}{4uv}, \quad y = \dfrac{w\big(u^2 + v^2\big)}{2uv}, \\
0 < u < v < \infty, \quad 0 < w < \infty;
\end{cases}
\end{gather*}
for $-t^2 + x^2 + y^2 < 0$
\begin{gather*}
\begin{cases}
{\rm d}s^2 = -{\rm d}u^2 + u^2\left(\dfrac{1}{v^2} + \dfrac{1}{w^2}\right)\big({\rm d}v^2 + {\rm d}w^2\big), \\
t+x = \dfrac{u}{vw}, \quad t-x = \dfrac{u\big(v^2 + w^2\big)^2}{4vw}, \quad y = \dfrac{u\big(w^2 - v^2\big)}{2vw}, \\
0 < u < \infty, \quad 0 < v < \infty, \quad 0 < w < \infty.
\end{cases}
\end{gather*}

\textbf{4.2.ii.} $A = J_{-1}(a^2) \oplus J_1(0) \oplus J_1(0)$, $a > 0$.
If $A$ has a multidimensional spacelike eigenspace corresponding to the smallest eigenvalue, we can cast~$A$ into this form using geometric equivalence. Hence, $A = a^2 e_0 \otimes e_0^{\flat}$ for some timelike unit vector~$e_0$. $L$ is reducible, and Algorithm~\ref{WP} gives the following warped product (letting $e_1$ be a spacelike unit vector orthogonal to~$e_0$):
\begin{align*}
\psi \colon \ N_0 \times_\rho \mathbb{S}^1 & \rightarrow \mathbb{E}^3_1, \\
(t e_0 + \tilde{x} e_1, p) &\mapsto t e_0 + \tilde{x}p,
\end{align*}
where $N_0 = \big\{te_0 + \tilde{x}e_1 \in \mathbb{E}^3_1 \, | \, \tilde{x} > 0 \big\}$ and $\rho(te_0 + \tilde{x}e_1) = \tilde{x}$. Following Algorithm~\ref{WP}, the restriction of $L$ to $N_0$ induces elliptic coordinates of type~II on~$N_0$, upon identifying $\operatorname{span}\{e_0, e_1\}$ with $\mathbb{E}^2_1$. Choosing the standard coordinate on~$\mathbb{S}^1$, we thus obtain the following web.

\textit{$23$. Elliptic-circular web II} (timelike rotational web IV),

\noindent for $|t| - \sqrt{x^2 + y^2} > a$
\begin{gather*}
\begin{cases}
{\rm d}s^2 = a^2\big(\cosh^2{v} - \cosh^2{u}\big)\big({\rm d}u^2 - {\rm d}v^2\big) + a^2 \sinh^2{v} \sinh^2{u} {\rm d}w^2, \\
 t = a \cosh{u} \cosh{v}, \quad x = a \sinh{v} \sinh{u} \cos{w}, \quad y = a \sinh{v} \sinh{u} \sin{w}, \\
0< u < v < \infty, \quad 0 < w < 2\pi, \quad a > 0,
\end{cases}
\end{gather*}
for $|t| - \sqrt{x^2 + y^2} < - a$
\begin{gather*}
\begin{cases}
{\rm d}s^2 = a^2\big(\cosh^2{u} - \cosh^2{v}\big)\big({\rm d}u^2 - {\rm d}v^2\big) + a^2 \cosh^2{v} \cosh^2{u} {\rm d}w^2, \\
t = a \sinh{v} \sinh{u}, \quad x = a \cosh{u} \cosh{v} \cos{w}, \quad y = a \cosh{u} \cosh{v} \sin{w},\\
0< v < u < \infty, \quad 0 < w < 2\pi, \quad a > 0;
\end{cases}
\end{gather*}
for $|t| + \sqrt{x^2 + y^2} < a$
\begin{gather*}
\begin{cases}
{\rm d}s^2 = a^2\big(\cos^2{u} - \cos^2{v}\big)\big({\rm d}u^2 - {\rm d}v^2\big) + a^2 \sin^2{u} \sin^2{v} {\rm d}w^2, \\
t = a \cos{u} \cos{v}, \quad x = a \sin{v} \sin{u} \cos{w}, \quad y = a \sin{v} \sin{u} \sin{w},\\
0< v < u < \frac{\pi}{2}, \quad 0 < w < 2\pi, \quad a>0,
\end{cases}
\end{gather*}

\textbf{4.2.iii.} $A = J_{-1}\big({-}a^2\big) \oplus J_1(0) \oplus J_1(0)$, $a > 0$.
If $A$ has a multidimensional spacelike eigen\-space corresponding to the largest eigenvalue, we may cast~$A$ into the above form using geometric equivalence. Hence, $A = -a^2 e_0 \otimes e_0^{\flat}$ for some timelike unit vector~$e_0$. $L$ is reducible, and Algorithm~\ref{WP} gives precisely the same warped product as in the previous case. In this case, however, the restriction of~$L$ to~$N_0$ induces elliptic coordinates of type~I on~$N_0$.

\textit{$24.$ Elliptic-circular web I} (timelike rotational web III)
\begin{gather*}
\begin{cases}
{\rm d}s^2 = a^2\big(\cosh^2{u} + \sinh^2{v}\big)\big({\rm d}u^2 - {\rm d}v^2\big) + a^2 \cosh^2{v} \sinh^2{u} {\rm d}w^2, \\
t = a \cosh{u} \sinh{v}, \quad x = a \cosh{v} \sinh{u} \cos{w}, \quad y = a \cosh{v} \sinh{u} \sin{w},\\
0< u < \infty, \quad 0 < v < \infty, \quad 0 < w < 2\pi, \quad a>0.
\end{cases}
\end{gather*}

\textbf{4.2.iv.} $A = J_{-1}(0) \oplus J_1(0) \oplus J_1\big(a^2\big)$, $a > 0$.
If $A$ has a multidimensional Lorentzian eigenspace corresponding to its smallest eigenvalue, we may cast it into this canonical form; i.e., $A = a^2 e_2 \otimes e_2^{\flat}$ for some spacelike unit vector $e_2$. $L$ is reducible, and following Algorithm~\ref{WP}, we must choose a unit vector in the multidimensional eigenspace of~$A$. Since this vector can be spacelike or timelike, we obtain two warped products, corresponding to different regions.

For a warped product with a spacelike vector $e_1$, Algorithm~\ref{WP} gives
\begin{align*}
\psi \colon \ N_0 \times_\rho \mathrm{dS}_1 &\rightarrow \mathbb{E}^3_1, \\
(ye_2 + \tilde{x}e_1, p) &\mapsto ye_2 + \tilde{x}p,
\end{align*}
where $N_0 = \big\{ ye_2 + \tilde{x}e_1 \in \mathbb{E}^3_1 \, | \, \tilde{x} > 0 \big\}$ and $\rho(ye_2 + \tilde{x}e_1) = \tilde{x}$. The restriction of~$L$ to $N_0$ induces elliptic coordinates on~$N_0$, upon identifying $\operatorname{span}\{e_1, e_2\}$ with~$\mathbb{E}^2$. For a~warped product with a~timelike vector~$e_0$, Algorithm~\ref{WP} gives
\begin{align*}
\psi \colon \ N_0 \times_\rho \mathbb{H}^1 &\rightarrow \mathbb{E}^3_1, \\
(-\tilde{t}e_0 + xe_2, p) & \mapsto \tilde{p} + xe_2,
\end{align*}
where $N_0 = \big\{ {-}\tilde{t}e_0 + xe_2 \in \mathbb{E}^3_1 \, | \, \tilde{t} > 0 \big\}$ and $\rho(-\tilde{t}e_0 + xe_2) = \tilde{t}$. The restriction of $L$ to~$N_0$ induces elliptic coordinates of type~I on~$N_0$, upon identifying $\operatorname{span}\{e_0, e_2\}$ with~$\mathbb{E}^2_1$. So, choosing the standard coordinate on~$\mathrm{dS}_1$ (resp.~$\mathbb{H}^1$), we obtain the following web.

\textit{$25.$ Elliptic-hyperbolic web I} (spacelike rotational web III),

\noindent for $-t^2 + x^2 > 0$
\begin{gather*}
\begin{cases}
{\rm d}s^2 = - a^2 \cosh^2{v} \cos^2{w}{\rm d}u^2 + a^2\big(\cosh^2{v} - \cos^2{w}\big)\big({\rm d}v^2 + {\rm d}w^2\big), \\
t = a \sinh{u} \cosh{v} \cos{w}, \quad x = a \cosh{u} \cosh{v} \cos{w}, \quad y = a \sinh{v} \sin{w},\\
0 < u < \infty, \quad 0 < v < \infty, \quad 0 < w < \frac{\pi}{2}, \quad a>0;
\end{cases}
\end{gather*}
for $-t^2 + x^2 < 0$
\begin{gather*}
\begin{cases}
{\rm d}s^2 = a^2\big(\cosh^2{u} + \sinh^2{v}\big)\big({\rm d}u^2 - {\rm d}v^2\big) + a^2 \cosh^2{u} \sinh^2{v}{\rm d}w^2, \\
t = a \cosh{u} \sinh{v} \cosh{w}, \quad x = a \cosh{u} \sinh{v} \sinh{w}, \quad y = a \sinh{u} \cosh{v},\\
0 < u < \infty, \quad 0 < v < \infty, \quad 0 < w < \infty, \quad a>0.
\end{cases}
\end{gather*}

\textbf{4.2.v.} $A = J_{-1}(0) \oplus J_1(0) \oplus J_1\big({-}a^2\big)$, $a > 0$.
If $A$ has a multidimensional Lorentzian eigenspace corresponding to the largest eigenvalue, we may cast $A$ into the above form, i.e., $A = -a^2 e_2 \otimes e_2^{\flat}$. $L$ is reducible, and Algorithm~\ref{WP} gives precisely the same warped products as in the previous case. In this case, however, the restriction of $L$ to $N_0$ induces elliptic coordinates in the first case, and elliptic coordinates of type II in the second. We thus obtain at the following web.

\textit{$26.$ Elliptic-hyperbolic web II}, (spacelike rotational web IV),

\noindent for $-t^2 + x^2 > 0$
\begin{gather*}
\begin{cases}
{\rm d}s^2 = - a^2 \sinh^2{v} \sin^2{w}{\rm d}u^2 + a^2\big(\cosh^2{v} - \cos^2{w}\big)\big({\rm d}v^2 + {\rm d}w^2\big), \\
t = a \sinh{u} \sinh{v} \sin{w}, \quad x = a \cosh{u} \sinh{v} \sin{w}, \quad y = a \cosh{v} \cos{w},\\
0 < u < \infty, \quad 0 < v < \infty, \quad 0 < w < \pi, \quad a>0;
\end{cases}
\end{gather*}
for $-t^2 + x^2 < 0$, $\sqrt{t^2 - x^2} - |y| > a$
\begin{gather*}
\begin{cases}
{\rm d}s^2 = a^2\big(\cosh^2{v} - \cosh^2{u}\big)\big({\rm d}u^2 - {\rm d}v^2\big) + a^2 \cosh^2{u} \cosh^2{v}{\rm d}w^2, \\
t = a \cosh{u} \cosh{v} \cosh{w}, \quad x = a \cosh{u} \cosh{v} \sinh{w}, \quad y = a \sinh{u} \sinh{v},\\
0 < u < v < \infty, \quad 0 < w < \infty, \quad a>0;
\end{cases}
\end{gather*}
for $-t^2 + x^2 < 0$, $\sqrt{t^2 - x^2} - |y| < - a$
\begin{gather*}
\begin{cases}
{\rm d}s^2 = a^2\big(\cosh^2{u} - \cosh^2{v}\big)\big({\rm d}u^2 - {\rm d}v^2\big) + a^2 \sinh^2{u} \sinh^2{v}{\rm d}w^2, \\
t = a \sinh{u} \sinh{v} \cosh{w}, \quad x = a \sinh{u} \sinh{v} \sinh{w}, \quad y = a \cosh{u} \cosh{v},\\
0 < v < u < \infty, \quad 0 < w < \infty, \quad a>0;
\end{cases}
\end{gather*}
for $-t^2 + x^2 < 0$, $\sqrt{t^2 - x^2} + |y| < a$
\begin{gather*}
\begin{cases}
{\rm d}s^2 = a^2\big(\cos^2{u} - \cos^2{v}\big)\big({\rm d}u^2 - {\rm d}v^2\big) + a^2 \cos^2{u} \cos^2{v}{\rm d}w^2, \\
t = a \cos{u} \cos{v} \cosh{w}, \quad x = a \cos{u} \cos{v} \sinh{w}, \quad y = a \sin{u} \sin{v},\\
0 < v < u < \frac{\pi}{2}, \quad 0 < w < \infty, \quad a>0.
\end{cases}
\end{gather*}

\textbf{4.2.vi.} $A = J_{2}(0)^{\rm T} \oplus J_1(0)$.
We now consider the above canonical form, i.e., $A = k \otimes k^{\flat}$ for some nonzero lightlike vector~$k$. Then~$A$ has a degenerate two-dimensional eigenspace, and so~$L$ is reducible. If~$k'$ is a null vector such that $\langle k, k' \rangle = 1$, Algorithm~\ref{WP} gives the following null warped product which decomposes~$L$
\begin{align*}
\psi \colon \ N_0 \times_\rho\mathbb{E}^1& \rightarrow \mathbb{E}^3_1, \\
(\tilde{\eta} k + \xi k', p)& \mapsto \xi (k' + p - \frac{1}{2}p^2k) + \tilde{\eta} k,
\end{align*}
where $N_0 = \big\{ \tilde{\eta}k + \xi k' \in \mathbb{E}^3_1 \, | \, \xi > 0 \big\}$ and $\rho(\tilde{\eta}k + \xi k') = \xi$. The restriction of~$L$ to~$N_0$ induces null elliptic coordinates of type~II on~$N_0$, upon identifying $\operatorname{span}\{k, k'\}$ with~$\mathbb{E}^2_1$. Thus, choosing the standard coordinate~$w$ on~$\mathbb{E}^1$ and rewriting in terms of Cartesian coordinates, we obtain the following web.

\textit{$27.$ Parabolically-embedded null elliptic web II} (null rotational web III),

\noindent for $-t^2 + x^2 + y^2 > |t - x|$
\begin{gather*}
\begin{cases}
{\rm d}s^2 = \big({\rm e}^{2u} - {\rm e}^{2v}\big)\big({\rm d}u^2 - {\rm d}v^2\big) + {\rm e}^{2(u+v)}{\rm d}w^2, \\
t - x = - {\rm e}^{u+v}, \quad t + x = 2\cosh{(u - v)} - w^2 {\rm e}^{u+v}, \quad y = w{\rm e}^{u + v},\\
-\infty < v < u < \infty, \quad -\infty < w < \infty;
\end{cases}
\end{gather*}
for $-t^2 + x^2 + y^2 < -|t - x|$
\begin{gather*}
\begin{cases}
{\rm d}s^2 = \big({\rm e}^{2v} - {\rm e}^{2u}\big)\big({\rm d}u^2 - {\rm d}v^2\big) + {\rm e}^{2(u+v)}{\rm d}w^2, \\
t - x = {\rm e}^{u+v}, \quad t + x = 2\cosh{(u - v)} + w^2 {\rm e}^{u+v}, \quad y = w{\rm e}^{u + v},\\
-\infty < u < v < \infty, \quad -\infty < w < \infty.
\end{cases}
\end{gather*}

\textbf{4.2.vii.} $A = J_{-2}(0)^{\rm T} \oplus J_1(0)$.
Consider now the above canonical form, i.e., $A = -k \otimes k^{\flat}$ for some nonzero null vector~$k$. Then~$A$ again has a degenerate two-dimensional eigenspace, $L$~is once again reducible, and Algorithm~\ref{WP} gives precisely the same warped product as in the previous case. In this case, however, the restriction of~$L$ to~$N_0$ yields null elliptic coordinates of type~I on~$N_0$. We therefore obtain the following web:

\textit{$28.$ Parabolically-embedded null elliptic web I} (null rotational web II)
\begin{gather*}
\begin{cases}
{\rm d}s^2 = \big({\rm e}^{2u} + {\rm e}^{2v}\big)\big({\rm d}u^2 - {\rm d}v^2\big) + {\rm e}^{2(u+v)}{\rm d}w^2, \\
t - x= {\rm e}^{u+v}, \quad t + x = 2\sinh{(u - v)} + w^2 {\rm e}^{u+v}, \quad y = w{\rm e}^{u + v}, \\
-\infty < u < \infty, \quad -\infty < v < \infty, \quad -\infty < w < \infty.
\end{cases}
\end{gather*}

\textbf{4.2.viii.} $A = J_{-1}(0) \oplus J_1(a) \oplus J_1(b)$, $0 < a < b$.
If $A$ is diagonalizable with real distinct eigenvalues, and has a timelike eigenvector corresponding to the smallest eigenvalue, then modulo geometric equivalence, $A$ takes the above form in some Cartesian coordinates $(t,x,y)$. $L$~is a~central irreducible CT in canonical form, and in these coordinates, $A$ and~$g$ take the form~\eqref{Ag} with $k=0$; therefore, equation~\eqref{centralICTa} is superfluous and the subspace $U$ referred to therein is trivial. So we need only equations~\eqref{centralICTb} and~\eqref{p(z)} to obtain the transformation between Cartesian coordinates $(t,x,y)$ and the separable coordinates $(u,v,w)$ induced by $L$.

The characteristic polynomial of $A$ is given by $B(z) = z(z-a)(z-b)$. The eigenvalues~$0$,~$a$ and~$b$ of~$A$ correspond to $t$, $x$ and $y$ respectively, and writing the characteristic polynomial of~$L$ as $p(z) = (z - u)(z - v)(z - w)$, we obtain the transformation equations via equation~\eqref{centralICTb}. Moreover, the metric coefficients in canonical coordinates are readily obtained from \eqref{ICTmetric}. The metric and transformation equations for all other irreducible CTs below follow just as straightforwardly by an application of equations \eqref{Ag}--\eqref{ICTmetric} as appropriate.

Therefore, in this case, we see the above irreducible CT induces the following web.

\textit{$29.$ Ellipsoidal web I} (asymmetric web IX) [B.1.a]
\begin{gather*}
\begin{cases}
{\rm d}s^2 = \dfrac{(u-v)(u-w)}{4u(u-a)(u-b)}{\rm d}u^2 + \dfrac{(v-u)(v-w)}{4v(v-a)(v-b)}{\rm d}v^2 + \dfrac{(w-u)(w-v)}{4w(w-a)(w - b)}{\rm d}w^2, \vspace{1mm}\\
t^2 = -\dfrac{uvw}{ab}, \quad x^2 = \dfrac{(a - u)(a - v)(a - w)}{a(b - a)}, \quad y^2 = -\dfrac{(b - u)(b - v)(b - w)}{b(b - a)}.
\end{cases}
\end{gather*}
We may find the ranges of the coordinates by imposing the constraints that the metric have Lorentzian signature, and that the Cartesian coordinates are real. We thus have, assuming $w < v < u$ without loss of generality,
\[ w < 0 < a < v < b < u \qquad \textrm{($w$ timelike)}.
\]

In order to find the coordinate domains one would need to analyze the discriminant~$\Delta$ of the characteristic polynomial of~$L$, which is here a polynomial of degree $8$ in $t$, $x$ and $y$. The regions where $\Delta > 0$ give the coordinate domains. While this task is generally intractable, we can deduce from the coordinate ranges that, for this web, there is only one adapted coordinate chart up to isometry. For a much more detailed exposition on the coordinate domains and singularities, we refer the reader to~\cite{Hinterleitner1998}.

At this point we would like to mention that in this web, and in all of the irreducible webs to follow, it is possible to set one of the parameters (in this case $a$ or $b$) equal to $1$ via a homothetic transformation of~$\mathbb{E}^3_1$, reducing the number of parameters. Indeed, this is what is typically done in~\cite{Hinterleitner1998}. However, under isometries, the above parameters (or more precisely, the number of parameters) are essential and cannot be removed.

\textbf{4.2.ix.} $A = J_{-1}(a) \oplus J_1(b) \oplus J_1(0)$, $0 < a < b$.
In the case that $A$ has the above canonical form modulo geometric equivalence, $L$ is irreducible and we readily obtain the transformation from Cartesian coordinates and the form of the metric from equations \eqref{Ag}--\eqref{ICTmetric}.

\textit{$30.$ Ellipsoidal web II} (not in \cite{Horwood2008a}) [B.1.d]
\begin{gather*}
\begin{cases}
{\rm d}s^2 = \dfrac{(u-v)(u-w)}{4u(u-a)(u-b)}{\rm d}u^2 + \dfrac{(v-u)(v-w)}{4v(v-a)(v-b)}{\rm d}v^2 + \dfrac{(w-u)(w-v)}{4w(w-a)(w - b)} {\rm d}w^2, \vspace{1mm} \\
t^2 = -\dfrac{(a - u)(a - v)(a - w)}{a(b - a)}, \quad x^2 = -\dfrac{(b - u)(b - v)(b - w)}{b(b - a)}, \quad y^2 = \dfrac{uvw}{a b}.
\end{cases}
\end{gather*}
We find the coordinate ranges by requiring that the Cartesian coordinates are real-valued, and that the metric is Lorentzian. Taking, without loss of generality, $w < v < u$, we have that the only possible coordinate ranges are the following
\begin{gather*}
w < v < 0 < a < b < u \qquad \textrm{($w$ timelike)},\\
0 < w < v < a < b < u \qquad \textrm{($v$ timelike)},\\
0 < a < w < v < b < u \qquad \textrm{($w$ timelike)},\\
0 < a < b < w < v < u \qquad \textrm{($v$ timelike)}.
\end{gather*}
Just as in the previous case, we cannot solve for the coordinate domains explicitly, but we can deduce from the above coordinate ranges that this web has four isometrically inequivalent regions, each of which realizes one of the above possibilities for the coordinate ranges.

\textbf{4.2.x.} $A = J_{-1}(b) \oplus J_1(a) \oplus J_1(0)$, $0 < a < b$.
If $A$ takes the above canonical form up to geometric equivalence, then $L$ is irreducible and we readily obtain the transformation equations from Cartesian coordinates, as well as the form of the metric in canonical coordinates from equations \eqref{Ag}--\eqref{ICTmetric}.

\textit{$31.$ Ellipsoidal web III} (not in \cite{Horwood2008a}) [B.1.c]
\begin{gather*}
\begin{cases}
{\rm d}s^2 = \dfrac{(u-v)(u-w)}{4u(u-b)(u-a)}{\rm d}u^2 + \dfrac{(v-u)(v-w)}{4v(v-b)(v-a)}{\rm d}v^2 + \dfrac{(w-u)(w-v)}{4w(w-b)(w - a)}{\rm d}w^2, \vspace{1mm} \\
t^2 = \dfrac{(b - u)(b - v)(b - w)}{b(b - a)}, \quad x^2 = \dfrac{(a - u)(a - v)(a - w)}{a(b - a)}, \quad y^2 = \dfrac{uvw}{a b}.
\end{cases}
\end{gather*}
Imposing the usual constraints, and taking $w < v < u$ without loss of generality, we find the only possibilities for the coordinates ranges are the following:
\begin{gather*}
w < v < 0 < u < a < b \qquad \textrm{($w$ timelike)},\\
0 < w < v < u < a < b \qquad \textrm{($v$ timelike)},\\
0 < w < a < v < u < b \qquad \textrm{($u$ timelike)},\\
0 < w < a < b < v < u \qquad \textrm{($v$ timelike)}.
\end{gather*}
This web has four isometrically inequivalent regions, each of which realizes one of the above possibilities for the coordinate ranges.

\textbf{4.2.xi.} $A = J_{1}(ib) \oplus J_1(-ib) \oplus J_1(c)$, $ b > 0$.
If $A$ has (non-real) complex eigenvalues, then using geometric equivalence, we may assume~$A$ takes the above form in complex-Cartesian coordinates $(z,\bar{z},y)$. Then, using equations \eqref{centralICTb} and \eqref{ICTmetric}, we obtain the form of the metric, as well as the transformation equations between the canonical coordinates and the complex-Cartesian coordinates. One obtains the equations for standard Cartesian coordinates using the transformation
\[ z = \frac{1}{\sqrt{2}}(x - {\rm i}t)\]
and observing that $\operatorname{Re}\big(z^2\big) = \frac{1}{2}\big(x^2 - t^2\big)$ and $|z|^2 = \frac{1}{2}\big(x^2 + t^2\big)$. This yields the following web.

\textit{$32.$ Complex ellipsoidal web} (asymmetric web X) [B.1.f]
\begin{gather*}
\begin{cases}
{\rm d}s^2 = \dfrac{(u-v)(u-w)}{4\big(u^2 + b^2\big)(u-c)}{\rm d}u^2 + \dfrac{(v-u)(v-w)}{4\big(v^2 + b^2\big)(v-c)}{\rm d}v^2 + \dfrac{(w-u)(w-v)}{4\big(w^2+b^2\big)(w - c)}{\rm d}w^2, \vspace{1mm}\\
x^2 - t^2 = \dfrac{b^2(u + v + w - c) + c(uv + uw +vw) - uvw}{b^2 + c^2}, \vspace{1mm}\\
x^2 + t^2 = \dfrac{\sqrt{u^2 + b^2}\sqrt{v^2 + b^2}\sqrt{w^2 + b^2}}{b\sqrt{b^2 + c^2}}, \quad y^2 = -\dfrac{(c-u)(c-v)(c-w)}{c^2 + b^2}.
\end{cases}
\end{gather*}
Imposing the usual constraints, and taking $w < v < u$ without loss of generality, we find the only possibilities for the coordinates ranges are the following:
\begin{gather*}
c < w < v < u \qquad \textrm{($v$ timelike)},\\
w < v < c < u \qquad \textrm{($w$ timelike)}.
\end{gather*}

\textbf{4.2.xii.} $A = J_{2}(0)^{\rm T} \oplus J_1(c)$, $c > 0$.
In this case, since $A$ does not have a multidimensional eigenspace, $L$ is an irreducible CT. So, using equations \eqref{centralICTa}--\eqref{centralICTb} and~(\ref{ICTmetric}), we obtain the form of the metric, and the transformation equations between canonical coordinates and the lightcone coordinates $(\eta, \xi, y)$ in which~$A$ takes the above form. Rewriting in terms of Cartesian coordinates, we have the following web.

\textit{$33.$ Null ellipsoidal web I} (asymmetric web VIII) [D.1.d]
\begin{gather*}
\begin{cases}
{\rm d}s^2 = \dfrac{(u-v)(u-w)}{4u^2(u-c)}{\rm d}u^2 + \dfrac{(v-u)(v-w)}{4v^2(v-c)}{\rm d}v^2 + \dfrac{(w-u)(w-v)}{4w^2(w - c)}{\rm d}w^2, \vspace{1mm} \\
(x + t)^2 = -\dfrac{uvw}{c}, \quad x^2 - t^2 = \dfrac{1}{c}(uv+uw+vw) - \dfrac{1}{c^2}uvw,\vspace{1mm}\\
y^2 = -\dfrac{(c-u)(c-v)(c-w)}{c^2}.
\end{cases}
\end{gather*}
Imposing the usual constraints, and taking $w < v < u$, we have
\[
w < 0 < v < c < u \qquad \textrm{($w$ timelike)}.
\]

\textbf{4.2.xiii.} $A = J_{2}(0)^{\rm T} \oplus J_1(-c)$, $c > 0$.
In this case, $L$ is irreducible and we readily obtain the form of the metric and the transformation equations between the canonical coordinates and the lightcone coordinates $(\eta, \xi, y)$ in which $A$ takes the above form. Rewriting in terms of Cartesian coordinates, we have the following web.

\textit{$34.$ Null ellipsoidal web II} (not in \cite{Horwood2008a}) [not in \cite{Hinterleitner1998}]
\begin{gather*}
\begin{cases}
{\rm d}s^2 = \dfrac{(u-v)(u-w)}{4u^2(u+c)}{\rm d}u^2 + \dfrac{(v-u)(v-w)}{4v^2(v+c)}{\rm d}v^2 + \dfrac{(w-u)(w-v)}{4w^2(w + c)}{\rm d}w^2, \vspace{1mm}\\
(x+t)^2 = \dfrac{uvw}{c}, \quad x^2 - t^2 = -\dfrac{1}{c}(uv+uw+vw) - \dfrac{1}{c^2}uvw,\vspace{1mm}\\
y^2 = \dfrac{(c+u)(c+v)(c+w)}{c^2}.
\end{cases}
\end{gather*}
Imposing the usual constraints, and taking $w < v < u$, we have
\begin{gather*}
-c < 0 < w < v < u \qquad \textrm{($v$ timelike)},\\
-c < w < v < 0 < u \qquad \textrm{($v$ timelike)}.
\end{gather*}

\textbf{4.2.xiv.} $A = J_{-2}(0)^{\rm T} \oplus J_1(c)$, $c > 0$.
In this case, $L$ is irreducible and we readily obtain the form of the metric and the transformation equations between the canonical coordinates and the lightcone coordinates $(\eta, -\xi, y)$ in which~$A$ takes the above form. Rewriting in terms of Cartesian coordinates, we have the following web.

\textit{$35.$ Null ellipsoidal web III} (not in~\cite{Horwood2008a}) [D.1.a]
\begin{gather*}
\begin{cases}
{\rm d}s^2 = \dfrac{(u-v)(u-w)}{4u^2(u-c)}{\rm d}u^2 + \dfrac{(v-u)(v-w)}{4v^2(v-c)}{\rm d}v^2 + \dfrac{(w-u)(w-v)}{4w^2(w - c)}{\rm d}w^2, \vspace{1mm}\\
(x+t)^2 = \dfrac{uvw}{c}, \quad x^2 - t^2 = \dfrac{1}{c}(uv+uw+vw) - \dfrac{1}{c^2}uvw,\vspace{1mm} \\
y^2 = -\dfrac{(c-u)(c-v)(c-w)}{c^2}.
\end{cases}
\end{gather*}
Imposing the usual constraints, and taking $w < v < u$, we have
\begin{gather*}
0 < c < w < v < u \qquad \textrm{($v$ timelike)},\\
0 < w < v < c < u \qquad \textrm{($w$ timelike)},\\
w < v < 0 < c < u \qquad \textrm{($w$ timelike)}.
\end{gather*}

\textbf{4.2.xv.} $A = J_{-2}(0)^{\rm T} \oplus J_1(-c)$, $ c > 0$.
In this case, $L$ is irreducible and we readily obtain the form of the metric and the transformation equations between the canonical coordinates and the lightcone coordinates $(\eta, -\xi, y)$ in which $A$ takes the above form. Rewriting in terms of Cartesian coordinates, we have the following web.

\textit{$36.$ Null ellipsoidal web IV} (not in \cite{Horwood2008a}) [D.1.b]
\begin{gather*}
\begin{cases}
{\rm d}s^2 = \dfrac{(u-v)(u-w)}{4u^2(u+c)}{\rm d}u^2 + \dfrac{(v-u)(v-w)}{4v^2(v+c)}{\rm d}v^2 + \dfrac{(w-u)(w-v)}{4w^2(w + c)}{\rm d}w^2, \vspace{1mm}\\
(x+t)^2 = -\dfrac{uvw}{c}, \quad x^2 - t^2 = -\dfrac{1}{c}(uv+uw+vw) - \dfrac{1}{c^2}uvw, \vspace{1mm}\\
y^2 = \dfrac{(c+u)(c+v)(c+w)}{c^2}.
\end{cases}
\end{gather*}
Imposing the usual constraints, and taking $w < v < u$, we have
\begin{gather*}
-c < w < 0 < v < u \qquad \textrm{($v$ timelike)},\\
-c < w < v < u < 0 \qquad \textrm{($v$ timelike)},\\
w < v <-c < u < 0 \qquad \textrm{($w$ timelike)}.
\end{gather*}

\textbf{4.2.xvi.} $A = J_{3}(0)^{\rm T}$.
If $A$ takes the above canonical form modulo geometric equivalence, then~$L$ is irreducible and we readily obtain the form of the metric, and the transformation equations between the canonical coordinates and the lightcone coordinates $(\eta, y, \xi)$ in which~$A$ takes the above form. Passing to the standard Cartesian coordinates, we have the following web.

\textit{$37.$ Null ellipsoidal web V} (asymmetric web VII) [F.1.a]
\begin{gather*}
\begin{cases}
{\rm d}s^2 = \dfrac{(u - v)(u - w)}{4u^3}{\rm d}u^2 + \dfrac{(v - u)(v - w)}{4v^3}{\rm d}v^2 + \dfrac{(u - w)(v - w)}{4w^3}{\rm d}w^2, \vspace{1mm} \\
(x+t)^2 = uvw, \quad (x + t) y = -\dfrac{1}{2}(uv + uw + vw), \quad - t^2 + x^2 + y^2 = u + v + w.
\end{cases}
\end{gather*}
Imposing the usual constraints, and taking $w < v < u$, we find
\begin{gather*}
0 < w < v < u \qquad \textrm{($v$ timelike)},\\
w < v < 0 < u \qquad \textrm{($w$ timelike)}.
\end{gather*}

\subsection[Non-null axial CTs, L=A+w otimes r\textasciicircum{}\{flat\} + r otimes w\textasciicircum{}\{flat\}, <w, w> not = 0]{Non-null axial CTs, $\boldsymbol{L = A + w \otimes r^{\flat} + r \otimes w^{\flat}}$, $\boldsymbol{\langle w, w \rangle \neq 0}$}

Since any non-null axial CT takes the above canonical form with $Aw = 0$, we need only classify the geometrically inequivalent forms for $A_c = A_{w^{\bot}}$; furthermore, in classifying canonical forms for non-null axial CTs, we may apply geometric equivalence in the subspace $w^{\bot}$, since in this case $\dim{\operatorname{span}\{w\}} = 1$ (cf.\ the remarks following Theorem~\ref{classificationtheorem}). For convenience, we choose coordinates such that $w = \partial_t$ when $w$ is timelike, and $w = \partial_y$ when $w$ is spacelike.

\textbf{4.3.i.} $\langle w, w \rangle = -1$, $A = 0$.
The above canonical form corresponds to any timelike axial CT for which $w^{\bot}$ is an eigenspace of~$A$. $L$ is then reducible and Algorithm~\ref{WP} gives the following warped product which decomposes $L$ (upon choosing a unit vector~$e_1$ in~$w^{\bot}$)
\begin{align*}
\begin{split}
\psi \colon \ N_0 \times_\rho \mathbb{S}^1 &\rightarrow \mathbb{E}^3_1, \\
(t w + \tilde{x} e_1, p) & \mapsto t w + \tilde{x} p,
\end{split}
\end{align*}
where $N_0 = \big\{ tw + \tilde{x}e_1 \in \mathbb{E}^3_1 \, | \, \tilde{x} > 0 \big\}$ and $\rho(tw + \tilde{x}e_1) = \tilde{x}$. The restriction of $L$ to $N_0$ induces timelike parabolic coordinates on $N_0$, upon identifying $\operatorname{span}\{w, e_1\}$ with $\mathbb{E}^2_1$. Choosing the standard coordinate on $\mathbb{S}^1$, we obtain the following web.

\textit{$38.$ Timelike parabolic-circular web} (timelike rotational web II)
\begin{gather*}
\begin{cases}
{\rm d}s^2 = \big(u^2 - v^2\big)\big({-}{\rm d}u^2 + {\rm d}v^2\big) + u^2 v^2{\rm d}w^2, \\
 t = \frac{1}{2}\big(u^2 + v^2\big), \quad x = uv \cos{w}, \quad y = uv \sin{w},\\
 0 < v < u < \infty, \quad 0 < w < 2\pi.
\end{cases}
\end{gather*}

\textbf{4.3.ii.} $\langle w, w \rangle = -1$, $A = J_{-1}(0) \oplus J_1(0) \oplus J_1(a)$, $a > 0$.
If $A_c$ is diagonalizable with real distinct eigenvalues, then we may set the least of them to $0$ by geometric equivalence in $w^{\bot}$. $L$~is irreducible, and using equations \eqref{axialICTa}--\eqref{ICTmetric}, we obtain the transformation equations between Cartesian coordinates and the induced separable coordinates, as well as the form of the metric in the separable coordinates.

\textit{$39.$ Timelike paraboloidal web} (asymmetric web IV) [C.a]
\begin{gather*}
\begin{cases}
{\rm d}s^2 = - \dfrac{(u-v)(u-w)}{4u(u - a)} {\rm d}u^2 + \dfrac{(u-v)(v-w)}{4v(v - a)} {\rm d}v^2 - \dfrac{(u - w)(v - w)}{4w(w - a)} {\rm d}w^2, \vspace{1mm}\\
t = - \dfrac{1}{2}(u + v + w),\quad x^2 = \dfrac{uvw}{a},\quad y^2 = - \dfrac{(u - a)(v - a)(w - a)}{a}.
\end{cases}
\end{gather*}
Imposing the usual constraints, and taking $w < v < u$ without loss of generality, we obtain the following admissible ranges for the above coordinates:
\begin{gather*}
w < v < 0 < u < a \qquad \textrm{($w$ timelike)},\\
0 < w < v < u < a \qquad \textrm{($v$ timelike)}.
\end{gather*}
While imposing our constraints also yields $0 < w < a < v < u$ as an admissible set of ranges, the corresponding coordinate patch is equivalent to the one induced by the first set of ranges above; this can be easily seen by multiplying~$L$ by~$-1$.

 \textbf{4.3.iii.} $\langle w, w \rangle = 1$, $ A = 0$.
The above canonical form corresponds to any spacelike axial CT for which $w^{\bot}$ is an eigenspace of~$A$. $L$~is then reducible, and following Algorithm~\ref{WP}, we must choose a unit vector in $w^{\bot}$. Since this vector can be spacelike or timelike, we obtain two warped products, depending on the point $\bar{p}$ with which we apply Algorithm~\ref{WP}.

For a spacelike vector $e_1$, Algorithm~\ref{WP} gives
\begin{align*}
\psi \colon \ N_0 \times_\rho \mathrm{dS}_1 &\rightarrow \mathbb{E}^3_1, \\
(yw + \tilde{x}e_1, p) &\mapsto t w + \tilde{x} p,
\end{align*}
where $N_0 = \big\{ yw + \tilde{x}e_1 \in \mathbb{E}^3_1 \, | \, \tilde{x} > 0 \big\}$ and $\rho(yw + \tilde{x}e_1) = \tilde{x}$. The restriction of~$L$ to~$N_0$ induces parabolic coordinates on $N_0$, upon identifying $\operatorname{span}\{e_1,w\}$ with $\mathbb{E}^2$. If instead we use a timelike vector~$e_0$, then Algorithm~\ref{WP} yields
\begin{align*}
\psi \colon \ N_0 \times_\rho \mathbb{H}^1 &\rightarrow \mathbb{E}^3_1, \\
(yw - \tilde{t}e_0, p) &\mapsto y w + \tilde{t} p,
\end{align*}
where $N_0 = \big\{ yw - \tilde{t}e_0 \in \mathbb{E}^3_1 \, | \, \tilde{t} > 0 \big\}$ and $\rho\big(yw - \tilde{t}e_0\big) = \tilde{t}$. The restriction of~$L$ to~$N_0$ induces spacelike parabolic coordinates on~$N_0$, upon identifying $\operatorname{span}\{e_0,w\}$ with~$\mathbb{E}^2_1$. So choosing the standard coordinate on~$\mathrm{dS}_1$ (resp.~$\mathbb{H}^1$), we obtain the following web.

\textit{$40.$ Spacelike parabolic-hyperbolic web} (spacelike rotational web II),\\
for $-t^2 + x^2 + y^2 > 0$, $-t^2 + x^2 > 0$
\begin{gather*}
\begin{cases}
{\rm d}s^2 = -v^2 w^2{\rm d}u^2 + \big(v^2 + w^2\big)\big({\rm d}v^2 + {\rm d}w^2\big), \\
t = vw \sinh{u}, \quad x = vw \cosh{u}, \quad y = \frac{1}{2}\big(v^2 - w^2\big), \\
-\infty < u < \infty, \quad 0 < v < \infty, \quad 0 < w < \infty;
\end{cases}
\end{gather*}
for $-t^2 + x^2 + y^2 > 0$, $-t^2 + x^2 < 0$
\begin{gather*}
\begin{cases}
{\rm d}s^2 = \big(u^2 - v^2\big)\big({\rm d}u^2 - {\rm d}v^2\big) + u^2 v^2{\rm d}w^2, \\
t = uv \cosh{w}, \quad x = uv \sinh{w}, \quad y = \frac{1}{2}\big(u^2 + v^2\big), \\
0 < v < u < \infty, \quad -\infty < w < \infty.
\end{cases}
\end{gather*}

\textbf{4.3.iv.} $\langle w, w \rangle = 1$, $ A = J_{-1}(a) \oplus J_1(0) \oplus J_1(0)$, $a > 0$.
If $A_c$ is diagonalizable with real distinct eigenvalues, then modulo geometric equivalence in~$w^{\bot}$, $A$~has the above form in some Cartesian coordinates. So~$L$ is an irreducible CT and we readily obtain the form of the metric as well as the transformation equations.

\textit{$41.$ Spacelike paraboloidal web} (asymmetric web IV) [C.b]
\begin{gather*}
\begin{cases}
{\rm d}s^2 = \dfrac{(u - v)(u - w)}{4u(u - a)}{\rm d}u^2 + \dfrac{(v - u)(v - w)}{4v(v - a)}{\rm d}v^2 + \dfrac{(w - u)(w - v)}{4w(w - a)}{\rm d}w^2, \vspace{1mm} \\
t^2 = -\dfrac{(u - a)(v - a)(w - a)}{a}, \quad x^2 = -\dfrac{uvw}{a}, \quad y = \dfrac{1}{2}(u + v + w).
\end{cases}
\end{gather*}
Imposing the usual constraints, and taking $w < v < u$ without loss of generality, we obtain the following admissible ranges for the above coordinates:
\begin{gather*}
w < 0 < a < v < u \qquad \textrm{($v$ timelike)},\\
w < 0 < v < u < a \qquad \textrm{($u$ timelike)},\\
w < v < u < 0 < a \qquad \textrm{($v$ timelike)}.
\end{gather*}

\textbf{4.3.v.} $\langle w, w \rangle = 1$, $A = J_{1}({\rm i}b) \oplus J_1(-{\rm i}b) \oplus J_1(0)$, $b > 0$.
If $A_c$ has (non-real) complex eigenvalues, then by geometric equivalence, we may assume $A$ takes the above form in complex-Cartesian coordinates $(z,\bar{z},y)$. Since $L$ is an irreducible CT, we readily obtain the form of the metric as well as the transformation equations between the separable coordinates and $(z,\bar{z},y)$. We pass to standard Cartesian coordinates $(t,x,y)$ precisely as described in~4.2.xi.

\textit{$42.$ Spacelike complex-paraboloidal web} (asymmetric web VI) [C.d]
\begin{gather*}
\begin{cases}
{\rm d}s^2 = \dfrac{(u - v)(u - w)}{4\big(u^2 + b^2\big)}{\rm d}u^2 + \dfrac{(v - u)(v - w)}{4\big(v^2 + b^2\big)}{\rm d}v^2 + \dfrac{(w - u)(w - v)}{4\big(w^2 + b^2\big)}{\rm d}w^2, \vspace{1mm} \\
t^2 - x^2 = uv + uw + vw - b^2,\quad y = \dfrac{1}{2}(u + v + w),\vspace{1mm}\\
t^2 + x^2 = \dfrac{\sqrt{u^2 + b^2}\sqrt{v^2 + b^2}\sqrt{w^2 + b^2}}{b}.
\end{cases}
\end{gather*}
Assuming as usual that $w < v < u$, we see that this gives a unique set of coordinate ranges, and the timelike coordinate is $v$.

\textbf{4.3.vi.} $\langle w, w \rangle = 1$, $ A = J_2(0)^{\rm T} \oplus J_1(0)$.
We consider now the above canonical form, i.e., $A_c = k \otimes k^{\flat}$ for some nonzero null vector $k \in w^{\bot}$. Note that this canonical form is equivalent to the one with $A_c = -k \otimes k^{\flat}$, which can be easily seen by multiplying~$L$ by~$-1$. Now, since $L$ is irreducible, using equations \eqref{axialICTa}--\eqref{ICTmetric}, we obtain the metric and the transformation equations from the separable coordinates to the lightcone coordinates in which~$A$ takes the above form. Passing to standard Cartesian coordinates, we have the following web.

\textit{$43.$ Spacelike null-paraboloidal web} (asymmetric web III) [F.1.c]
\begin{gather*}
\begin{cases}
{\rm d}s^2 = \dfrac{(u - v)(u - w)}{4u^2}{\rm d}u^2 - \dfrac{(u - v)(v - w)}{4v^2}{\rm d}v^2 + \dfrac{(v - w)(u - w)}{4w^2}{\rm d}w^2,\vspace{1mm} \\
(t+x)^2 = uvw, \quad t^2 - x^2 = uv + uw + vw, \quad y = \dfrac{1}{2}(u + v + w).
\end{cases}
\end{gather*}
Imposing the usual constraints, and taking $w < v < u$ without loss of generality, we obtain the following admissible ranges for the above coordinates:
\begin{gather*}
0 < w < v < u \qquad \textrm{($v$ timelike)},\\
w < v < 0 < u \qquad \textrm{($v$ timelike)}.
\end{gather*}

\subsection[Null axial CTs, L=A+ w otimes r\textasciicircum{}\{flat\}+r otimes w\textasciicircum{}\{flat\}, w not= 0, <w,w>=0]{Null axial CTs, $\boldsymbol{L = A + w \otimes r^{\flat} + r \otimes w^{\flat}}$, $\boldsymbol{w \neq 0}$, $\boldsymbol{\langle w, w \rangle = 0}$}

Any null axial CT can be put into the above form using geometric equivalence, with the canonical form for $A$ given by Theorem~\ref{classificationtheorem}. Recall from the remarks following Theorem~\ref{classificationtheorem} that~$D$ is the $A$-invariant subspace $\operatorname{span}\big\{w,Aw,A^2w, \dots \big\}$. In classifying canonical forms for null axial CTs, we may apply geometric equivalence in the subspace $D^{\bot}$ insofar that $\dim{D} \leq 2$ (this is only relevant for~4.4.i below). For convenience, we choose coordinates such that $w = \partial_\eta = \partial_t + \partial_x$.

\textbf{4.4.i.} $w \neq 0$, $\langle w, w \rangle = 0$, $A = J_{2}(0)^{\rm T} \oplus J_1(0)$.
We consider the above canonical form, i.e., $A = k \otimes k^{\flat}$ for some nonzero null vector $k$ such that $\langle w, k \rangle = 1$. This case is equivalent to the null axial CT with $A = - k \otimes k ^{\flat}$. Since $L$ is an irreducible CT, using equations \eqref{axialICTa}--\eqref{ICTmetric}, we obtain the metric and the transformation equations from the separable coordinates to the lightcone coordinates in which $A$ takes the above form. Passing to standard Cartesian coordinates, we have the following web.

\textit{$44.$ Null paraboloidal web I} (asymmetric web II) [E.1.a]
\begin{gather*}
\begin{cases}
{\rm d}s^2 = \dfrac{(u - v)(u - w)}{4u}{\rm d}u^2 - \dfrac{(u - v)(v - w)}{4v}{\rm d}v^2 + \dfrac{(v - w)(u - w)}{4w}{\rm d}w^2, \vspace{1mm}\\
x + t = \dfrac{1}{8}\big(u^2 + v^2 + w^2\big) - \dfrac{1}{4}(uv + uw + vw),\quad x - t = u + v + w,\quad y^2 = uvw.
\end{cases}
\end{gather*}
Imposing the usual constraints, and taking $w < v < u$ without loss of generality, we obtain the following admissible ranges for the above coordinates:
\begin{gather*}
0 < w < v < u \qquad \textrm{($v$ timelike)},\\
w < v < 0 < u \qquad \textrm{($w$ timelike)}.
\end{gather*}

\textbf{4.4.ii.} $w \neq 0$, $\langle w, w \rangle = 0$, $A = J_{3}(0)^{\rm T}$.
We finally consider the above canonical form for a null axial CT~$L$, which is easily seen to be irreducible. We readily obtain the metric and the transformation equations between the separable coordinates and the lightcone coordinates in which~$A$ takes the above form. Passing to standard Cartesian coordinates, we obtain the following final web.

\textit{$45.$ Null paraboloidal web II} (asymmetric web I) [G]
\begin{gather*}
\begin{cases}
{\rm d}s^2 = \dfrac{1}{4}(u - v)(u - w){\rm d}u^2 - \dfrac{1}{4}(u - v)(v - w){\rm d}v^2 + \dfrac{1}{4}(u - w)(v - w){\rm d}w^2,\vspace{1mm} \\
x + t = \dfrac{1}{16}(u-v-w)(u+v-w)(u-v+w),\quad x - t = u + v + w,\vspace{1mm} \\
y = \dfrac{1}{8}\big(u^2 - v^2 -w^2\big) - \dfrac{1}{4}(uv + uw - vw).
\end{cases}
\end{gather*}
Assuming as usual that $w < v < u$, we see that this gives a unique set of coordinate ranges, and the timelike coordinate is~$v$.

\section{Concluding remarks}\label{sec5}

The above analysis illustrates the ease with which the method of concircular tensors yields an invariant classification of the orthogonal separable webs in $\mathbb{E}^3_1$. This approach appears to be much less computationally demanding than other methods described in the introduction, and is in many ways much more elementary, since the crux of this method is essentially a problem of linear algebra. Furthermore, provided we have determined all possible canonical forms modulo geometric equivalence, the theory presented in~\cite{Rajaratnam2014} and~\cite{Rajaratnam2014b} guarantees that the above 45 webs are all the orthogonal separable webs in $\mathbb{E}^3_1$; and insofar that our analysis of each case is complete, we can be confident that the 88 inequivalent adapted coordinate charts determined above are exhaustive.\looseness=-1

It is worth noting that the use of concircular tensors in classifying separable webs is likely to be fruitful only in spaces of constant curvature, since in spaces of non-constant curvature the space of concircular tensors will in general be very small.

In~\cite{Hinterleitner1998}, one finds 31 inequivalent coordinate charts corresponding to irreducible webs; moreover, it is asserted therein that there are 51 inequivalent coordinate systems arising from reducible webs, based on computations done in~\cite{Hinterleitner1998}. Counting up the results from Section~\ref{sec4}, we find 33 inequivalent charts from the irreducible webs, and 55 inequivalent charts from the reducible webs. As indicated above, the only two coordinate charts not appearing in \cite{Hinterleitner1998} are those from null ellipsoidal web II (web number 34 above). The discrepancy in the number of reducible chart domains seems to be a result of overlooked equivalences. For instance, in \cite{Hinterleitner1998} it appears that the Rindler web (web number 5 above and coordinate system~B.6.b in~\cite{Hinterleitner1998}) is counted as inducing only one inequivalent chart domain instead of two.

Any other discrepancies between the equations appearing here, and those appearing in~\cite{Hinterleitner1998} or~\cite{Horwood2008a} can be attributed to the various freedoms in defining the separable coordinates (i.e., geometric equivalence and reparametrizations of the web. See also the remarks following 4.2.viii above).

The method presented here achieves simultaneously both a global classification of the orthogonal separable webs, as well as the equations for the separable coordinates adapted to each web; the former gives the intrinsic geometric object of interest, while the latter gives the information necessary to orthogonally separate the Hamilton--Jacobi and Klein--Gordon equations given in the introduction.

\appendix

\section{Self-adjoint operators in Minkowski space}\label{appA}
In this appendix, we review some of the key results regarding self-adjoint operators on $n$-dimensional Minkowski space $\mathbb{E}^n_1$, whose theory and classification differ tremendously from the Euclidean case. We simply quote the main results in this section, and refer the reader to \cite{Rajaratnam2014} for details and proofs.

We first define a \textit{$k$-dimensional Jordan block} with eigenvalue $\lambda$, $J_k(\lambda)$, and a \textit{$k$-dimensional skew-normal matrix} $S_k$, to be the following $k \times k$ matrices:
\begin{gather*}
 J_k(\lambda) :=
	 \begin{pmatrix}
		\lambda & 1 & & 0 & \\
		& \lambda & \ddots & & \\
		& & \ddots & 1 & \\
		& & & \lambda & 1 \\
		& 0 & & & \lambda
 	\end{pmatrix}, \qquad S_k := \begin{pmatrix}
	 	0 & & & & 1 \\
	 & & & 1 & \\
		 & & \iddots & & \\
		 & 1 & & & \\
	 	1 & & & & 0
 \end{pmatrix}.
\end{gather*}
A sequence of vectors in which the metric (restricted to their span) takes the form $\varepsilon S_k$ is called a \textit{skew-normal sequence}. Recall that a linear operator $A\colon \mathbb{E}^n_\nu \rightarrow \mathbb{E}^n_\nu$ is self-adjoint with respect to the scalar product if $\langle Ax, y \rangle = \langle x, Ay \rangle$ for all~$x$ and~$y$. This holds if and only if the contravariant or covariant tensor metrically equivalent to~$A$ is symmetric. Since the metric is not positive definite in $\mathbb{E}^n_1$, our classification of self-adjoint operators will specify the forms taken by both~$A$ \textit{and}~$g$ in an appropriate basis. The canonical form for the pair~$(A,g)$ is called the \textit{metric-canonical form} or \textit{metric-Jordan form} for~$A$.

For this purpose, we introduce a signed integer $\varepsilon k \in \mathbb{Z}$, where $\varepsilon = \pm 1$ and $k \in \mathbb{N}$, and write $A = J_{\varepsilon k}(\lambda)$ as a shorthand for the pair $A = J_k(\lambda)$ and $g = \varepsilon S_k$. For square matrices $A_1$ and $A_2$, we also define the block diagonal matrix
\begin{gather*}
 A_1 \oplus A_2 :=
	 \begin{pmatrix}
		A_1 & 0 \\
		0 & A_2
 	\end{pmatrix}.
\end{gather*}
We write $J_{\varepsilon k}(\lambda) \oplus J_{\delta m}(\mu)$ as a shorthand the pair $J_k(\lambda) \oplus J_m(\mu)$ and $g = \varepsilon S_k \oplus \delta S_m$. We now summarize the different possible canonical forms for a self-adjoint operator $A$ in $\mathbb{E}^n_1$: 

 \textbf{Case 1.} $A$ is diagonalizable with real eigenvalues. In this case, there is a basis such that
\begin{gather*}
 A = J_{-1}(\lambda_1) \oplus J_1(\lambda_2) \oplus \cdots \oplus J_1(\lambda_n).
\end{gather*}
Equivalently, $A$ is diagonalized in Cartesian coordinates.

\textbf{Case 2.} $A$ has a complex eigenvalue $\lambda = a + {\rm i}b$ with $b \neq 0$. Since $A$ is real, $\bar{\lambda}$ must be another eigenvalue; in Minkowski space, all other eigenvalues must be real. Then,
\begin{gather*}
 A = J_1(\lambda) \oplus J_1(\bar{\lambda}) \oplus J_1(\lambda_3) \oplus \dots \oplus J_1(\lambda_n)
\end{gather*}
in some orthogonal basis where the first two vectors are complex. Notice that since they are complex, we may assume they have length squared~$+1$.

\textbf{Case 3.} $A$ has real eigenvalues but is not diagonalizable. Then there are three possibilities for the metric-canonical form. The first two occur when
\begin{gather*}
 A = J_{\varepsilon2}(\lambda) \oplus J_1(\lambda_3) \oplus \dots \oplus J_1(\lambda_n)
\end{gather*}
with $\varepsilon = \pm 1$, in some basis where the first two vectors are null. The last case occurs when
\begin{gather*}
 A = J_3(\lambda) \oplus J_1(\lambda_4) \oplus \dots \oplus J_1(\lambda_n)
\end{gather*}
in some basis where the first and third vectors are null; the second is spacelike. Notice that in Minkowski space, a metric-Jordan block $J_{-3}(\lambda)$ is inadmissible. These are all the possibilities for the canonical forms of self-adjoint endomorphisms in Minkowski space.

\vspace{-2mm}

\subsection*{Acknowledgements}

The authors wish to thank 
K.~Rajaratnam and the anonymous referees for their careful reading of the paper and a number of helpful suggestions and comments. We also wish to acknowledge financial support from the Natural Sciences and Engineering Research Council of Canada in the form of a Undergraduate Student Research Award (CV) and a Discovery Grant~(RGM).

\vspace{-2mm}

\pdfbookmark[1]{References}{ref}
\LastPageEnding


\begin{thebibliography}{99}
\footnotesize\itemsep=-1.0pt

\bibitem{Benenti2005b}
Benenti S., Special symmetric two-tensors, equivalent dynamical systems,
 cofactor and bi-cofactor systems, \href{https://doi.org/10.1007/s10440-005-1138-9}{\textit{Acta Appl. Math.}} \textbf{87}
 (2005), 33--91.

\bibitem{Benenti2002a}
Benenti S., Chanu C., Rastelli G., Remarks on the connection between the
 additive separation of the {H}amilton--{J}acobi equation and the
 multiplicative separation of the {S}chr\"odinger equation. {I}.~{T}he
 completeness and {R}obertson conditions, \href{https://doi.org/10.1063/1.1506180}{\textit{J.~Math. Phys.}} \textbf{43}
 (2002), 5183--5222.

\bibitem{Bocher1894}
B\^ocher M., \"Uber die Riehenentwickelungen der Potentialtheory, B.G.~Teubner,
 Leipzig, 1894.

\bibitem{Bruce2001}
Bruce A.T., McLenaghan R.G., Smirnov R.G., A geometrical approach to the
 problem of integrability of {H}amiltonian systems by separation of variables,
 \href{https://doi.org/10.1016/S0393-0440(01)00017-1}{\textit{J.~Geom. Phys.}} \textbf{39} (2001), 301--322.

\bibitem{Cochran2011}
Cochran C.M., McLenaghan R.G., Smirnov R.G., Equivalence problem for the
 orthogonal webs on the 3-sphere, \href{https://doi.org/10.1063/1.3578773}{\textit{J.~Math. Phys.}} \textbf{52} (2011),
 053509, 22~pages, \href{https://arxiv.org/abs/1009.4244}{arXiv:1009.4244}.

\bibitem{Cochran2017}
Cochran C.M., McLenaghan R.G., Smirnov R.G., Equivalence problem for the
 orthogonal separable webs in 3-dimensional hyperbolic space, \href{https://doi.org/10.1063/1.4983998}{\textit{J.~Math.
 Phys.}} \textbf{58} (2017), 063513, 43~pages.

\bibitem{Crampin2003}
Crampin M., Conformal {K}illing tensors with vanishing torsion and the
 separation of variables in the {H}amilton--{J}acobi equation,
 \href{https://doi.org/10.1016/S0926-2245(02)00140-7}{\textit{Differential Geom. Appl.}} \textbf{18} (2003), 87--102.

\bibitem{Crampin2007}
Crampin M., Concircular vector fields and special conformal {K}illing tensors,
 in Differential Geometric Me\-thods in Mechanics and Field Theory, Academia
 Press, Gent, 2007, 57--70.

\bibitem{Eisenhart1934}
Eisenhart L.P., Separable systems of {S}tackel, \href{https://doi.org/10.2307/1968433}{\textit{Ann. of Math.}}
 \textbf{35} (1934), 284--305.

\bibitem{Hinterleitner1996}
Hinterleitner F., Examples of separating coordinates for the {K}lein--{G}ordon
 equation in {$(2+1)$}-dimensional flat space-time, \href{https://doi.org/10.1063/1.531552}{\textit{J.~Math. Phys.}}
 \textbf{37} (1996), 3032--3040.

\bibitem{Hinterleitner1998}
Hinterleitner F., Global properties of orthogonal separable coordinates for the
 {K}lein--{G}ordon equation in {$(2+1)$}-dimensional flat space-time,
 \textit{\"Osterreich. Akad. Wiss. Math.-Natur. Kl. Sitzungsber.~II}
 \textbf{207} (1998), 133--171.

\bibitem{Horwood2008b}
Horwood J.T., On the theory of algebraic invariants of vector spaces of
 {K}illing tensors, \href{https://doi.org/10.1016/j.geomphys.2007.12.004}{\textit{J.~Geom. Phys.}} \textbf{58} (2008), 487--501.

\bibitem{Horwood2007}
Horwood J.T., McLenaghan R.G., Transformation to pseudo-{C}artesian coordinates
 in locally flat pseudo-{R}iemannian spaces, \href{https://doi.org/10.1016/j.geomphys.2006.12.001}{\textit{J.~Geom. Phys.}}
 \textbf{57} (2007), 1435--1440.

\bibitem{Horwood2008a}
Horwood J.T., McLenaghan R.G., Orthogonal separation of variables for the
 {H}amilton--{J}acobi and wave equations in three-dimensional {M}inkowski
 space, \href{https://doi.org/10.1063/1.2823971}{\textit{J.~Math. Phys.}} \textbf{49} (2008), 023501, 48~pages,
 \href{https://arxiv.org/abs/#2}{arXiv:10.1063/1.2823971}.

\bibitem{Horwood2009}
Horwood J.T., McLenaghan R.G., Smirnov R.G., Hamilton--{J}acobi theory in
 three-dimensional {M}inkowski space via {C}artan geometry, \href{https://doi.org/10.1063/1.3094719}{\textit{J.~Math.
 Phys.}} \textbf{50} (2009), 053507, 41~pages.

\bibitem{Kalnins1975}
Kalnins E.G., On the separation of variables for the {L}aplace equation
 {$\Delta \Psi +K^{2}\Psi =0$} in two- and three-dimensional {M}inkowski
 space, \href{https://doi.org/10.1137/0506033}{\textit{SIAM~J. Math. Anal.}} \textbf{6} (1975), 340--374.

\bibitem{Kalnins1986b}
Kalnins E.G., Separation of variables for {R}iemannian spaces of constant
 curvature, \textit{Pitman Monographs and Surveys in Pure and Applied
 Mathematics}, Vol.~28, Longman Scientific \& Technical, Harlow, John Wiley \&
 Sons, Inc., New York, 1986.

\bibitem{Kalnins2018}
Kalnins E.G., Kress J.M., Miller Jr. W., Separation of variables and
 superintegrability. The symmetry of solvable systems, \textit{IOP Expanding Physics},
 \href{https://doi.org/10.1088/978-0-7503-1314-8}{IOP Publishing}, Bristol, 2018.

\bibitem{Kalnins1976a}
Kalnins E.G., Miller Jr. W., Lie theory and separation of variables.
 {IX}.~{O}rthogonal {$R$}-separable coordinate systems for the wave equation
 {$\psi_{tt}-\Delta^{2}\psi =0$}, \href{https://doi.org/10.1063/1.522900}{\textit{J.~Math. Phys.}} \textbf{17} (1976),
 331--355.

\bibitem{Kalnins1986}
Kalnins E.G., Miller Jr. W., Separation of variables on {$n$}-dimensional
 {R}iemannian manifolds. {I}. {T}he {$n$}-sphere {$S_n$} and {E}uclidean
 {$n$}-space {${\bf R}^n$}, \href{https://doi.org/10.1063/1.527088}{\textit{J.~Math. Phys.}} \textbf{27} (1986),
 1721--1736.

\bibitem{McLenaghan2020}
McLenaghan R.G., Rastelli G., Valero C., Complete separability of the
 {H}amilton--{J}acobi equation for the charged particle orbits in a
 {L}i\'enard--{W}iechert field, \href{https://doi.org/10.1063/5.0030305}{\textit{J.~Math. Phys.}} \textbf{61} (2020),
 122903, 29~pages.

\bibitem{McLenaghan2002}
McLenaghan R.G., Smirnov R.G., Intrinsic characterizations of orthogonal
 separability for natural {H}amiltonians with scalar potentials on
 pseudo-{R}iemannian spaces, \href{https://doi.org/10.2991/jnmp.2002.9.s1.12}{\textit{J.~Nonlinear Math. Phys.}} \textbf{9}
 (2002), suppl.~1, 140--151.

\bibitem{McLenaghan2004}
McLenaghan R.G., Smirnov R.G., The D., An extension of the classical theory of
 algebraic invariants to pseudo-{R}iemannian geometry and {H}amiltonian
 mechanics, \href{https://doi.org/10.1063/1.1644902}{\textit{J.~Math. Phys.}} \textbf{45} (2004), 1079--1120.

\bibitem{Olevsky1950}
Olevsky M.N., Triorthogonal systems in spaces of constant curvature in which
 the equation {$\Delta_2 u + \lambda u = 0$} admits complete separation of
 variables, \textit{Math. USSR. Sb.} \textbf{27} (1950), 379--427.

\bibitem{Rajaratnam2014}
Rajaratnam K., Orthogonal separation of the {H}amilton--{J}acobi equation on
 spaces of constant curvature, {M}aster's Thesis, University of Waterloo,
 2014, available at \url{http://hdl.handle.net/10012/8350}.

\bibitem{Rajaratnam2014b}
Rajaratnam K., McLenaghan R.G., Classification of {H}amilton--{J}acobi
 separation in orthogonal coordinates with diagonal curvature,
 \href{https://doi.org/10.1063/1.4893335}{\textit{J.~Math. Phys.}} \textbf{55} (2014), 083521, 16~pages,
 \href{https://arxiv.org/abs/1404.2565}{arXiv:1404.2565}.

\bibitem{Rajaratnam2014a}
Rajaratnam K., McLenaghan R.G., Killing tensors, warped products and the
 orthogonal separation of the {H}amilton--{J}acobi equation, \href{https://doi.org/10.1063/1.4861707}{\textit{J.~Math.
 Phys.}} \textbf{55} (2014), 013505, 27~pages, \href{https://arxiv.org/abs/1404.3161}{arXiv:1404.3161}.

\bibitem{Rajaratnam2016}
Rajaratnam K., McLenaghan R.G., Valero C., Orthogonal separation of the
 {H}amilton--{J}acobi equation on spaces of constant curvature, \href{https://doi.org/10.3842/SIGMA.2016.117}{\textit{SIGMA}}
 \textbf{12} (2016), 117, 30~pages, \href{https://arxiv.org/abs/1607.00712}{arXiv:1607.00712}.

\bibitem{Robertson1927}
Robertson H.P., Bemerkung \"uber separierbare {S}ysteme in der
 {W}ellenmechanik, \href{https://doi.org/10.1007/BF01451624}{\textit{Math. Ann.}} \textbf{98} (1928), 749--752.

\bibitem{Smirnov2004}
Smirnov R.G., Yue J., Covariants, joint invariants and the problem of
 equivalence in the invariant theory of {K}illing tensors defined in
 pseudo-{R}iemannian spaces of constant curvature, \href{https://doi.org/10.1063/1.1805728}{\textit{J.~Math. Phys.}}
 \textbf{45} (2004), 4141--4163, \href{https://arxiv.org/abs/math-ph/0407028}{arXiv:math-ph/0407028}.

\bibitem{Valero2019}
Valero C., McLenaghan R.G., Classification of the orthogonal separable webs for
 the {H}amilton--{J}acobi and {L}aplace--{B}eltrami equations on 3-dimensional
 hyperbolic and de {S}itter spaces, \href{https://doi.org/10.1063/1.5043066}{\textit{J.~Math. Phys.}} \textbf{60}
 (2019), 033501, 30~pages, \href{https://arxiv.org/abs/1811.04536}{arXiv:1811.04536}.

\end{thebibliography}
\end{document}